# Report on article:

# „P=NP Linear programming formulation of the Traveling Salesman Problem”[1]

Radosław Hofman, Poznań, 2006

## 1 Introduction

Document mentioned in title of this article was published in first version in 2004. In moment research was started version revised on 16.01.2006 was available and this article refers mainly to that version. At this moment newer version revised on 16.10.2006 is available, but main idea is same. This article contains formulation of TSP[2] problem. Then problem is modeled as Integer Linear Programming[3] and after this author claims that solution is equivalent to non-integer version of the same linear programming problem – non integer solution is presenting many solutions at once.

Example from article for graph with 6 nodes:

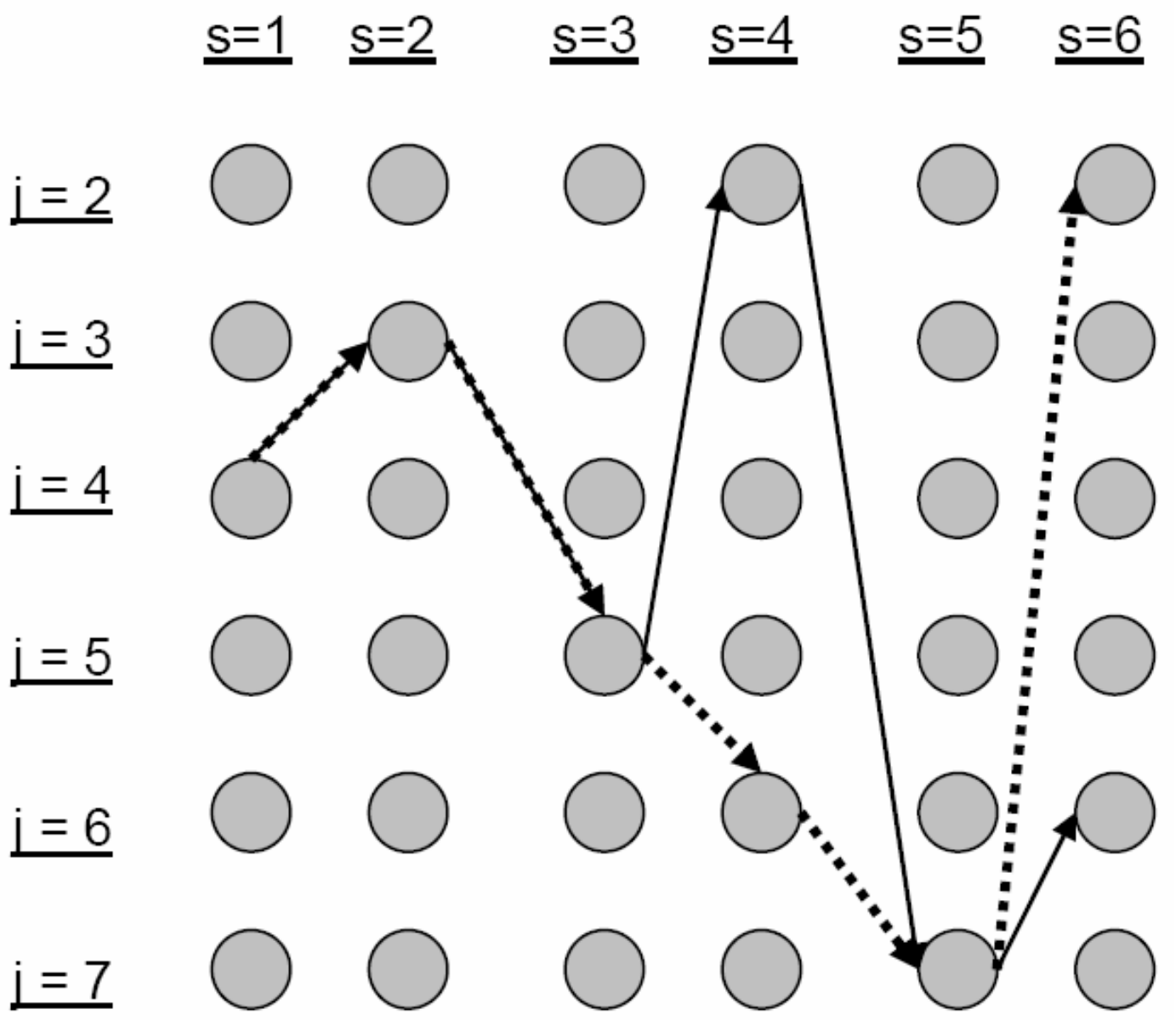

*Figure 1 Example of flow*

If solution proposed in article is correct then P=NP[4] with all consequences. Although solving instances of problem is more then difficult (it uses $n^9$ variables) it makes solution possible to be found in polynomially bounded time.

---

[1] Author of document: Moustapha Diaby, article is available under http://arxiv.org/ftp/cs/papers/0609/0609005.pdf or http://www.business.uconn.edu/users/mdiaby/TSPLP/Paper/Version_12_29_05/TSP_LP.pdf. New version is available under: http://www.business.uconn.edu/users/mdiaby/TSPLP/10_13_06_Rvn/Paper/tsplpPaper.pdf

[2] See http://en.wikipedia.org/wiki/Traveling_salesman_problem

[3] See http://en.wikipedia.org/wiki/Integer_programming

[4] See http://en.wikipedia.org/wiki/P_vs_NP

Article mentioned in title is logically equivalent to article[5] from same author about algorithm solving QAP[6] problem. Counter examples to discussed article can be used to prove incorrectness of second article as well. Recently there appeared article from Sergey Gubin using analogical idea – counter example fits for this as well.

This article shows why solution proposed in mentioned publication is not correct.

# 2 Close look at definitions

## 2.1 $x_{i,s,j}$ variables (first dimension)

First part of transformation is introducing variables $x_{i,s,j}$ (spanning over nodes and paths). Index of variable should be interpreted as:

- $i$ – begin node
- $j$ – end node
- $s$ – number of step

For $n$ nodes we obtain $n^3$ $x_{i,s,j}$ variables. Can TSP problem be expressed using only this variables? Of course if we know optimal solution then we may assign 1 to proper $x_{i,s,j}$ and 0 to other, but is it possible to state equations enabling computation of optimal solution?

Answer is positive. If we have condition that $x_{i,s,j}=\{0,1\}$ and we require:

- only one $x_{i,s,j}$ from beginning: $\sum_{i,j\in<1,n>} x_{i,1,j} = 1$
- sum of input flow is equal to output flow at each node: $\sum_{i\in<1,n>} x_{i,s,j} = \sum_{i\in<1,n>} x_{j,s+1,i}$ for all $j\in<1,n>$ and $s\in<1,n-1>$
- each node must be visited exactly once: $\sum_{i\in<1,n>,s\in<1,n>} x_{i,s,j} = 1$ for all $j\in<1,n>$
- there can not exist any flow from node to itself $\sum_{i,j\in<1,n>,s\in<1,n-1>} x_{i,s,j} * c_{i,j}$ for all $i\in<1,n>$

Every assignment compliant with above equations is feasible solution for corresponding TSP problem if then one looks for minimal sum of then optimal solution would be found. Of course this requires that all $x_{i,s,j}$ are included - solution proposed in article omits one $x_{i,s,j}$ (assuming that first flow is from node and it needs no variable).

What will happen when we relax requirement for $x_{i,s,j}=\{0,1\}$? Above restrictions can be expressed in less then:

- 1 equation
- $n^2$ equations
- $n$ equations
- $n$ equations

To assign values to $n^3$ variables. It is obvious then if solution is to be found it will be set with infinite number of solutions.

[5] See http://www.business.uconn.edu/users/mdiaby/QAPLP/Paper/Oct_1_06_Rvn/qaplpPaper.pdf
[6] See http://en.wikipedia.org/wiki/Quadratic_assignment_problem

For better understanding meaning of *x*'s we will show it in table for *n*=4:

| *i,j* | | *s* | | | |
|---|---|---|---|---|---|
| | | *1* | *2* | *3* | *4* |
| *1* | *1* | 0 | 0 | 0 | 0 |
| *1* | *2* | | | | |
| *1* | *3* | | | | |
| *1* | *4* | | | | |
| *2* | *1* | | | | |
| *2* | *2* | 0 | 0 | 0 | 0 |
| *2* | *3* | | | | |
| *2* | *4* | | | | |
| *3* | *1* | | | | |
| *3* | *2* | | | | |
| *3* | *3* | 0 | 0 | 0 | 0 |
| *3* | *4* | | | | |
| *4* | *1* | | | | |
| *4* | *2* | | | | |
| *4* | *3* | | | | |
| *4* | *4* | 0 | 0 | 0 | 0 |

*Table 1 Interpretation of* $x_{i,s,j}$

Every row represent flow from *i* to *j* and columns represent stages *s*. Restrictions mentioned at the beginning of this section can be understood as:

- sum of first column equals to 1
- sum of *s+1* column equals to sum of *s* column (especially for rows whose first index equals to second index in *s* column)
- sum of rows having equal first index is equal to 1
- every cell in row having equal first and second index is equal to 0

Now we will show method how to build counter example. It is important to get idea because verifying all variables in last section needs very long time (while idea is the same).

One may think that the method produces incorrect solution when optimal path does not contain minimum values (that is necessary condition for providing solution better then optimal for integer version of problem). Let us then consider graph consisting of four nodes A, B, C, D with weights:

| | A | B | C | D |
|---|---|---|---|---|
| **A** | 99 | 10 | 20 | 20 |
| **B** | 20 | 99 | 25 | 25 |
| **C** | 20 | 14 | 99 | 25 |
| **D** | 20 | 14 | 25 | 99 |

*Table 2 Costs for graph ABCD*

Without loosing generality we may assume that every path starts at node A – we have then 3! paths:

| Path | Cost |
|---|---|
| A→B→C→D→A | 80 |
| A→B→D→C→A | 80 |
| A→C→B→D→A | 79 |
| A→C→D→B→A | 79 |
| A→D→B→C→A | 79 |
| A→D→C→B→A | 79 |

*Table 3 Possible solutions for graph ABCD*

Now what we need is to preserve factor (that optimal solution does not include single arc) of this graph even after one (any) arc would be removed – such requirement follows meaning of $x_{i,s,j}$ – it represents part of universe from point of view of one arc in spanned graph. It is easy to do it by substituting every arc with two arcs (A: 1→2, B: 3→4, C: 5→6, D: 7→8) setting costs in such way that original costs are preserved and adding costs to prevent optimal paths "go" between nodes – in this case 15 is enough (to get between and get out would require double of this what is sufficient for such solution to be greater then optimal)):

| | | A | | B | | C | | D | |
|---|---|---|---|---|---|---|---|---|---|
| | | **1** | **2** | **3** | **4** | **5** | **6** | **7** | **8** |
| **A** | **1** | 99 | 5 | 15 | 15 | 15 | 15 | 15 | 15 |
| | **2** | 15 | 99 | 5 | 15 | 15 | 15 | 15 | 15 |
| **B** | **3** | 15 | 15 | 99 | 10 | 15 | 15 | 15 | 15 |
| | **4** | 10 | 15 | 15 | 99 | 15 | 15 | 15 | 15 |
| **C** | **5** | 15 | 15 | 15 | 15 | 99 | 8 | 15 | 15 |
| | **6** | 12 | 15 | 6 | 15 | 15 | 99 | 17 | 15 |
| **D** | **7** | 15 | 15 | 15 | 15 | 15 | 15 | 99 | 7 |
| | **8** | 13 | 15 | 7 | 15 | 18 | 15 | 15 | 99 |

*Table 4 Costs for graph 1..8*

Shortest solution in this graph are:

| Cost | Number of solutions | Example |
|---|---|---|
| 83 | 24 | 1→2→3→5→6→4→7→8→1 |
| 82 | 24 | 1→2→3→5→6→7→8→4→1 |
| 81 | 8 | 1→2→5→6→3→7→8→4→1 |
| 80 | 16 | 1→2→3→4→5→6→7→8→1 |
| 79 | 32 | 1→2→5→6→3→4→7→8→1 |

*Table 5 Best solutions for graph 1..8*

Now we set up linear programming problem and solve it with Soplex [Wun1996]. Optimal solution found is 75! With variables assigned (all other are zeros):

| Variable | Value |
|---|---|
| $x_{3,1,4}$ | 0,50 |
| $x_{6,1,5}$ | 0,25 |
| $x_{8,1,7}$ | 0,25 |
| $x_{4,2,1}$ | 0,50 |
| $x_{5,2,6}$ | 0,25 |
| $x_{7,2,8}$ | 0,25 |
| $x_{1,3,2}$ | 0,50 |
| $x_{6,3,5}$ | 0,25 |
| $x_{8,3,7}$ | 0,25 |
| $x_{2,4,3}$ | 0,50 |

| Variable | Value |
|---|---|
| $x_{5,4,6}$ | 0,25 |
| $x_{7,4,8}$ | 0,25 |
| $x_{3,5,4}$ | 0,50 |
| $x_{6,5,5}$ | 0,25 |
| $x_{8,5,7}$ | 0,25 |
| $x_{4,6,1}$ | 0,50 |
| $x_{5,6,6}$ | 0,25 |
| $x_{7,6,8}$ | 0,25 |
| $x_{1,7,2}$ | 0,50 |
| $x_{6,7,5}$ | 0,25 |
| $x_{8,7,7}$ | 0,25 |
| $x_{2,8,3}$ | 0,50 |
| $x_{5,8,6}$ | 0,25 |
| $x_{7,8,8}$ | 0,25 |

*Table 6 Non-integer assignment of* $x_{i,s,j}$

Graphical interpretation of this table is:

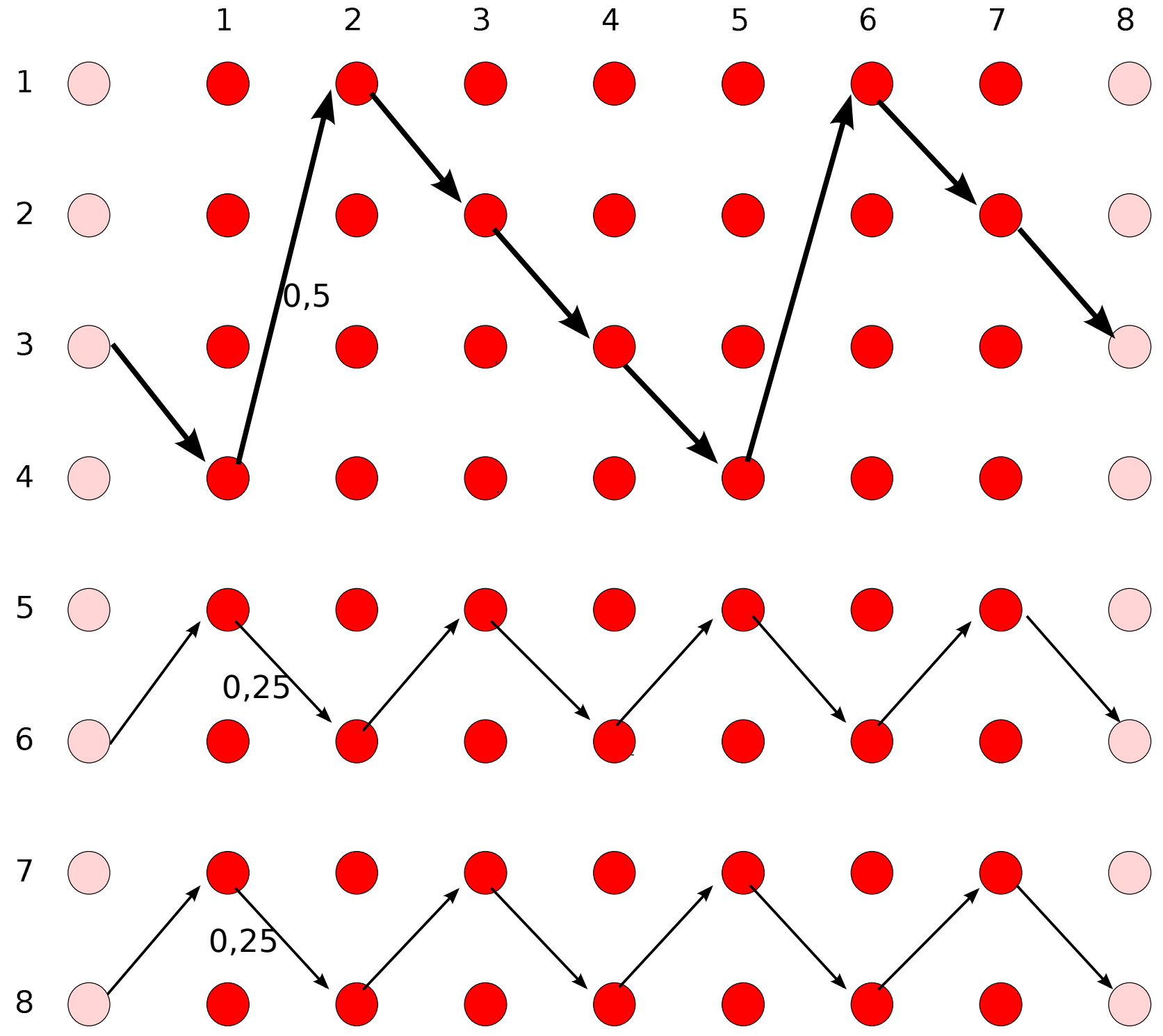

*Figure 2 Example of feasible flow for model* $x_{i,s,j}$

First layer was appended to figure in order to show flow at first stage – from layer zero to first. This layer is equivalent to last layer (8) because all paths form cycles in graph.

This proves that although solution was correct for integer version of problem it is incorrect after relaxing integer requirement. Why? This model after relaxing assumption that $x_{i,s,j}$ may be only 0 or 1 is cannot "react" on decision made in previous steps.

Important thing is to understand why solution is incorrect. Let us consider following model (let it be based on original TSP problem):

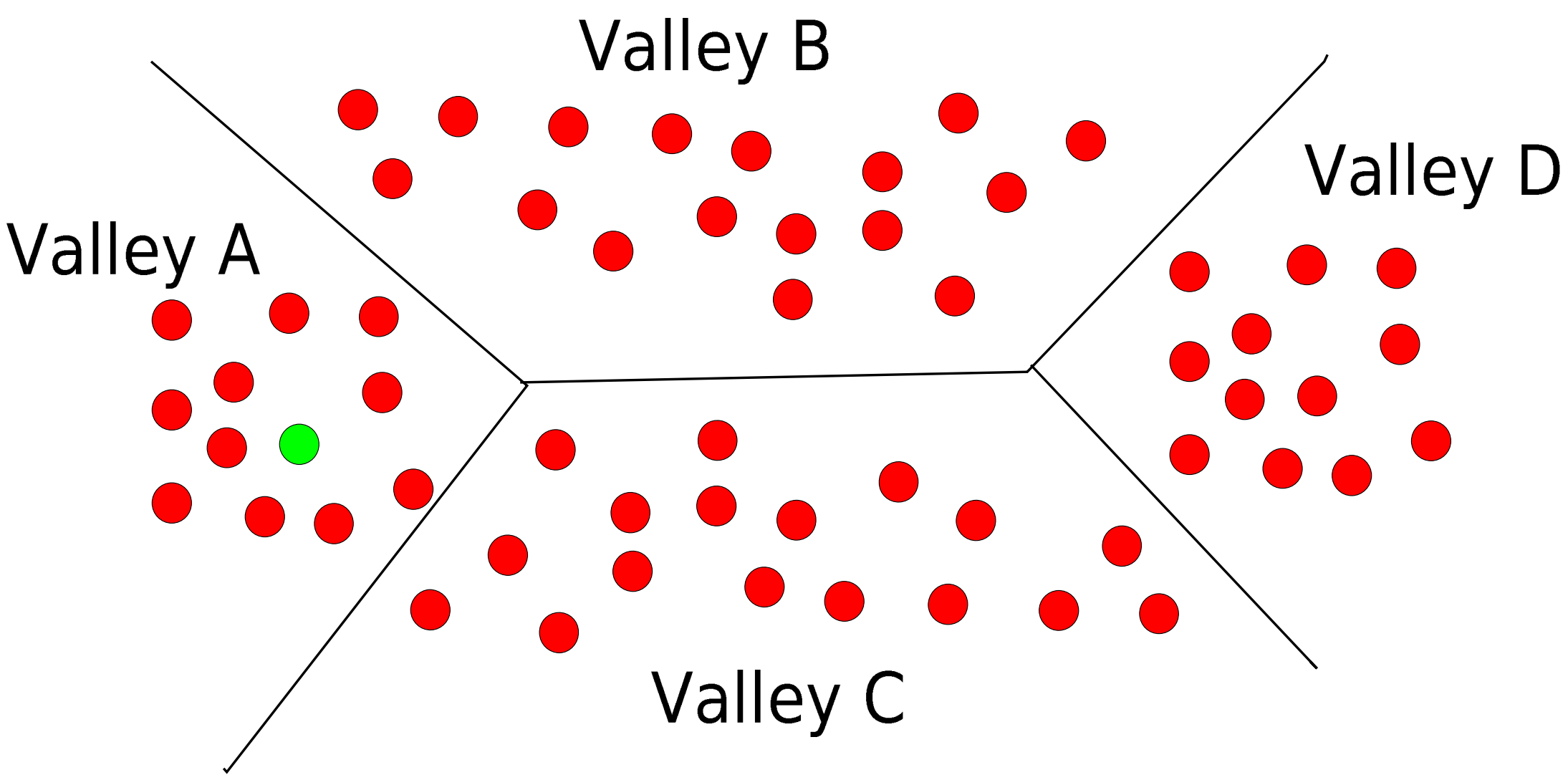

*Figure 3 Four valleys*

Imagine *v* valleys and all cities placed in this valleys. Cost of traveling from town to town in valley is very small (negligible), let us say equal to 1. When salesmen wants to go to other valley he has to pass mountain chain – this is of a huge cost, let us say equal to 1000. Salesman lives in one of towns in valley A. Optimal solution must then contain 4 mountain chains crossing (from A to B, B to C, C to D and D to A), its overall cost is then 4*1000+*X* (*X* is cost of in valley traveling and is very small). But after relaxing restriction on salesman being one person (in fact this is consequence of relaxing assumption of integer programming), then following solution is correct in model, has lower overall cost and is incorrect for TSP problem:

- start from town in valley A
- visit once all cities in valley A
- ½ of salesman travels to valley B and ½ of salesman travels to valley C – cost of this "move" is ½ * 2 * 1000 = 1000
- visit **twice** every city in valley B and C (this is done by proper "half" of salesman)
- "halfs" follow to valley D – again cost is ½ * 2 * 1000 = 1000
- visit all towns in valley D
- return to starting town – this move crosses mountains so it cost is 1000

We can clearly see that every city was visited by whole salesman (2 * ½ = 1 ☺) and total cost of this trip is 3*1000+$X_1$ (we saw that optimal solution was 4*1000+*X*). Now let us have a look on flow according to model:

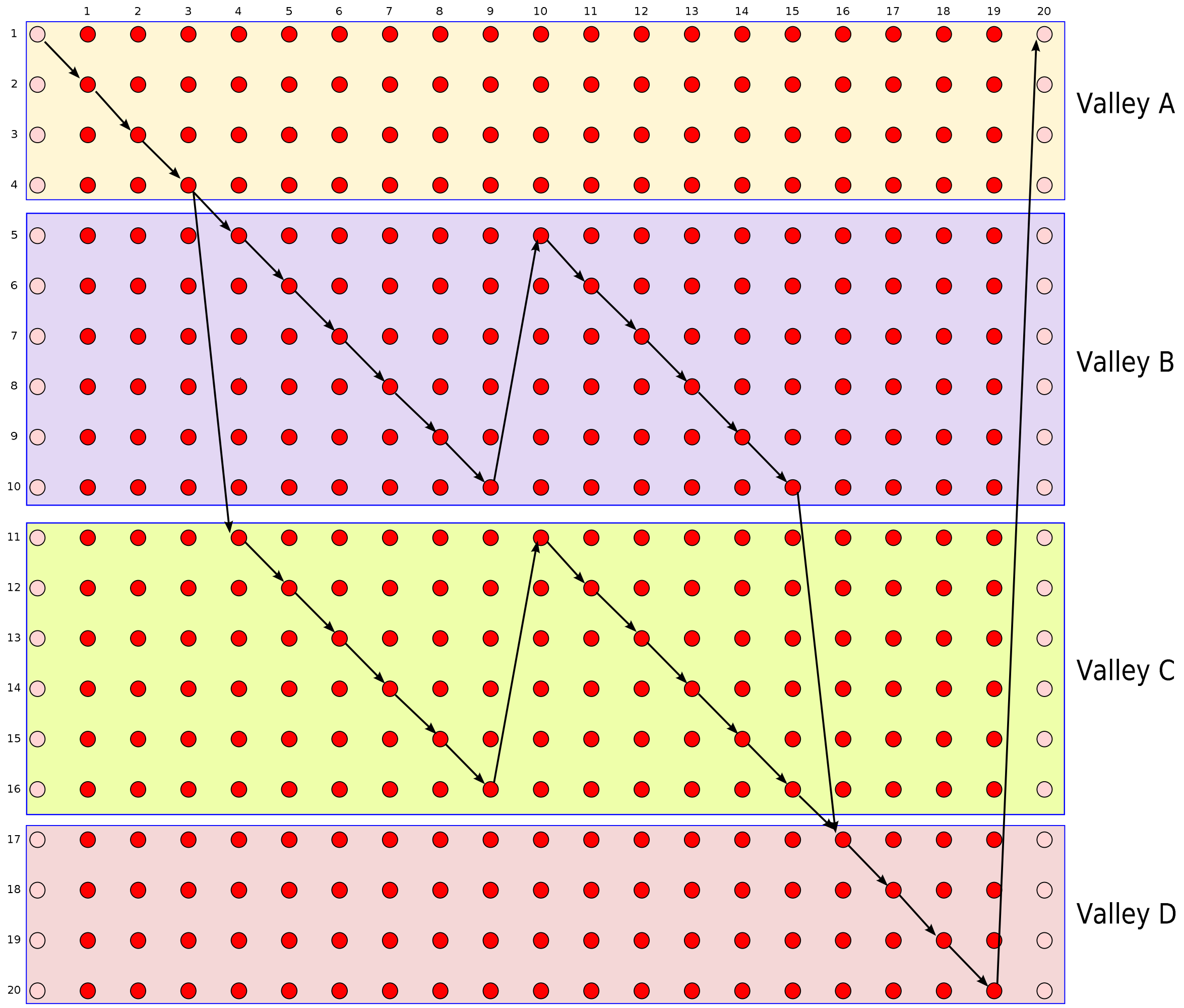

*Figure 4 Four valleys path on diagram*

Now let us consider if adding more dimensions of checking can correct situation.

## 2.2 $y_{i,s,j,u,p,v}$ variables (second dimension)

Author introduces variables $y_{i,s,j,u,p,v}$ which indexes can be interpreted as combination of two $x_{i,s,j}$: flow caused on $x_{u,p,v}$ by $x_{i,s,j}$ ($p>s$). Author allows usage of $x_{i,s,j}$ and $y_{i,s,j,u,p,v}$ introducing $y_{i,s,j,i,s,j}=x_{i,s,j}$.

How can it be imagined? If we look back at table 1 and imagine recursive replacement of each cell with table 1. We will obtain following table:

| i,j | | u,v | | s,r | | | | | | | | | | | | | | | |
|---|---|---|---|---|---|---|---|---|---|---|---|---|---|---|---|---|---|---|---|
| | | | | 1 | 1 | 1 | 1 | 2 | 2 | 2 | 2 | 3 | 3 | 3 | 3 | 4 | 4 | 4 | 4 |
| | | | | 1 | 2 | 3 | 4 | 1 | 2 | 3 | 4 | 1 | 2 | 3 | 4 | 1 | 2 | 3 | 4 |
| 1 | 1 | 1 | 1 | | | | | | | | | | | | | | | | |
| | | 1 | 2 | | | | | | | | | | | | | | | | |
| | | 1 | 3 | | | | | | | | | | | | | | | | |
| | | 1 | 4 | | | | | | | | | | | | | | | | |
| | | 2 | 1 | | | | | | | | | | | | | | | | |
| | | 2 | 2 | | | | | | | | | | | | | | | | |
| | | 2 | 3 | | | | | | | | | | | | | | | | |
| | | 2 | 4 | | | | | | | | | | | | | | | | |

| i,j | | u,v | | s,r | | | | | | | | | | | | | | | |
|---|---|---|---|---|---|---|---|---|---|---|---|---|---|---|---|---|---|---|---|
| | | | | 1 | 1 | 1 | 1 | 2 | 2 | 2 | 2 | 3 | 3 | 3 | 3 | 4 | 4 | 4 | 4 |
| | | | | 1 | 2 | 3 | 4 | 1 | 2 | 3 | 4 | 1 | 2 | 3 | 4 | 1 | 2 | 3 | 4 |
| | | 3 | 1 | | | | | | | | | | | | | | | | |
| | | 3 | 2 | | | | | | | | | | | | | | | | |
| | | 3 | 3 | | | | | | | | | | | | | | | | |
| | | 3 | 4 | | | | | | | | | | | | | | | | |
| | | 4 | 1 | | | | | | | | | | | | | | | | |
| | | 4 | 2 | | | | | | | | | | | | | | | | |
| | | 4 | 3 | | | | | | | | | | | | | | | | |
| | | 4 | 4 | | | | | | | | | | | | | | | | |
| 1 | 2 | 1 | 1 | | | | | | | | | | | | | | | | |
| | | 1 | 2 | | | | | | | | | | | | | | | | |
| | | 1 | 3 | | | | | | | | | | | | | | | | |
| | | 1 | 4 | | | | | | | | | | | | | | | | |
| | | 2 | 1 | | | | | | | | | | | | | | | | |
| | | 2 | 2 | | | | | | | | | | | | | | | | |
| | | 2 | 3 | | | | | | | | | | | | | | | | |
| | | 2 | 4 | | | | | | | | | | | | | | | | |
| | | 3 | 1 | | | | | | | | | | | | | | | | |
| | | 3 | 2 | | | | | | | | | | | | | | | | |
| | | 3 | 3 | | | | | | | | | | | | | | | | |
| | | 3 | 4 | | | | | | | | | | | | | | | | |
| | | 4 | 1 | | | | | | | | | | | | | | | | |
| | | 4 | 2 | | | | | | | | | | | | | | | | |
| | | 4 | 3 | | | | | | | | | | | | | | | | |
| | | 4 | 4 | | | | | | | | | | | | | | | | |
| … | | | | | | | | | | | | | | | | | | | |
| 4 | 4 | 1 | 1 | | | | | | | | | | | | | | | | |
| | | 1 | 2 | | | | | | | | | | | | | | | | |
| | | 1 | 3 | | | | | | | | | | | | | | | | |
| | | 1 | 4 | | | | | | | | | | | | | | | | |
| | | 2 | 1 | | | | | | | | | | | | | | | | |
| | | 2 | 2 | | | | | | | | | | | | | | | | |
| | | 2 | 3 | | | | | | | | | | | | | | | | |
| | | 2 | 4 | | | | | | | | | | | | | | | | |
| | | 3 | 1 | | | | | | | | | | | | | | | | |
| | | 3 | 2 | | | | | | | | | | | | | | | | |
| | | 3 | 3 | | | | | | | | | | | | | | | | |
| | | 3 | 4 | | | | | | | | | | | | | | | | |
| | | 4 | 1 | | | | | | | | | | | | | | | | |
| | | 4 | 2 | | | | | | | | | | | | | | | | |
| | | 4 | 3 | | | | | | | | | | | | | | | | |
| | | 4 | 4 | | | | | | | | | | | | | | | | |

*Table 7 Interpretation of* $y_{i,s,j,u,p,v}$

Different colors present replacement done on table 1 with nesting itself to one cell.

If integer model using only $x_{i,s,j}$ variables was sufficient to express one possible solution then after adding more variables (there are $O(n^6)$ variables now) it still should have enough power to do so. Now author requires such restrictions to variables:

- only one $x_{i,s,j}$ from beginning: $\sum_{i,j\in<1,n>} y_{i,1,j,i,1,j} = 1$ (equation 2.17 in old version of article, 2.6 in new version)
- sum of input flow is equal to output flow at each node: $\sum_{i\in<1,n>} y_{i,s,j,i,s,j} = \sum_{i\in<1,n>} y_{j,s+1,i,j,s+1,i}$ for all $j\in<1,n>$ and $s\in<1,n\text{-}1>$ (equation 2.18 in old version of article, skipped in new version)
- every flow must be caused by flow from first step $y_{i,s,j,i,s,j} = \sum_{u,v\in<1,n>} y_{u,1,v,i,s,j}$ for all $i,j\in<1,n>$ and $s\in<2,n\text{-}2>$ (equation 2.19 in old version article, 2.7 in new version)
- sum of flow caused to enter node must be equal to sum of flow leaving this node $\sum_{k\in<1,n>} y_{i,s,j,k,r,t} = \sum_{k\in<1,n>} y_{i,s,j,t,r+1,k}$ for all $i,j,t\in<1,n>$ and $s,r\in<1,n\text{-}3>$ (of course $s\leq r$) (equation 2.20 in old version of article, 2.8 in new version)
- flow in first step must "reach" every node with same value $y_{i,1,j,i,1,j} = \sum_{k\in<1,n>,r\in<2,n>} y_{i,1,j,k,r,t}$ for each $i,j,t\in<1,n>$ (of course $t\neq i$ and $t\neq j$) (equation 2.24 in old version of article, skipped in new version)
- flow cannot cause flow on previous layers $\sum_{k,t\in<1,n>,r\in<1,n>,r<s} y_{i,s,j,k,r,t} = 0$ for each $i,j\in<1,n>$ and $s\in<1,n\text{-}1>$ (equation 2.26 in old version of article, 2.14 in new version - part 1)
- flow cannot cause flow on same layer $\sum_{k,t\in<1,n>,(i,j)\neq(k,t)} y_{i,s,j,k,s,t} = 0$ for each $i,j\in<1,n>$ and $s\in<1,n\text{-}1>$ (equation 2.26 in old version of article, 2.14 in new version - part 2)
- flow cannot be "broken" $\sum_{k,t\in<1,n>,j\neq k} y_{i,s,j,k,s+1,t} = 0$ for each $i,j$ $<1,n>$ and $s\in<1,n>$ (equation 2.26 - part 3)
- each node can be visited only once (I) - flow cannot cause flow to the same node $\sum_{k\in<1,n>,r\in<1,n>,r>s} y_{i,s,j,k,r,i} = 0$ for each $i,j\in<1,n>$ and $s\in<1,n>$ (equation 2.26 in old version of article, 2.14 in new version - part 4)
- each node can be visited only once (II) - flow cannot cause flow to the same node $\sum_{k\in<1,n>,r\in<1,n>,r>s} y_{i,s,j,i,r,k} = 0$ for each $i,j\in<1,n>$ and $s\in<1,n>$ (equation 2.26 in old version of article, 2.14 in new version - part 5)
- each node can be visited only once (III) - flow cannot cause flow to the same node $\sum_{k\in<1,n>,r\in<1,n-1>,r>s} y_{i,s,j,k,r,j} = 0$ for each $i,j\in<1,n>$ and $s\in<1,n>$ (equation 2.26 in old version of article, 2.14 in new version - part 6)
- each node can be visited only once (IV) - flow cannot cause flow to the same node $\sum_{k\in<1,n>,r\in<1,n-1>,r>s+1} y_{i,s,j,j,r,k} = 0$ for each $i,j\in<1,n>$ and $s\in<1,n>$ (equation 2.26 in old version of article, 2.14 in new version - part 7)
- there can be no flow from node to itself (I) $\sum_{k,t\in<1,n>,r\in<1,n-1>} y_{i,s,i,k,r,t} = 0$ for each $i\in<1,n>$ and $s\in<1,n>$ (equation 2.26 in old version of article, 2.14 in new version - part 8)
- there can be no flow from node to itself (II) $\sum_{k,t\in<1,n>,r\in<1,n-1>} y_{k,r,t,i,s,i} = 0$ for each $i\in<1,n>$ and $s\in<1,n>$ (equation 2.26 in old version of article, 2.14 in new version - part 9)

Ranges are repeated after original old version of article but for example in 2.19 and 2.20 they should be larger to reach every $y_{i,s,j,u,p,v}$. Ranges should be extended if one relaxes assumption that flows start from node 1. We would have possible flow $y_{1,1,2,8,8,1}$.

As we can see author did not include requirement stated by me in previous section - sum of flow at each node must be equal to 1 ( $\sum_{j\in<1,n>,s\in<1,n>} y_{i,s,j,i,s,j} = 1$ for every $i\in<1,n>$); sum of flow caused by arc must be equal at every stage ( $\sum_{u,v\in<1,n>} y_{i,s,j,u,p,v} = y_{i,s,j,i,s,j}$ for every $i,j\in<1,n>$, $s\in<1,n-1>$ and $p\in<1,n>$) and sum of flow caused by arc must "reach" every node with same value. These requirements are covered by 2.25 in old version of article, 2.13 in new version but it is not implemented in BLP definition in new version. Without them even integer solution for $y$'s model may be incorrect! Of course adding this condition does not make non-integer solution be correct.

After implementation of this conditions and running solver we can see that this conditions are weaker then discussed in previous section. As it was mentioned we nest table 1 in every cell of this table but there are not stated all necessary conditions (for nested tables), what makes solution "seen" from perspective of one arc differ from solution which cost is given as result. Before providing result one have to add conditions for nested tables:

- every flow connection must relate to flow I $y_{i,s,j,i,s,j} = \sum_{u,v\in<1,n>} y_{i,s,j,u,p,v}$ for all $i,j\in<1,n>$ and $s\in<1,n-1>$ and $p\in<2,n>$ where $p>s$ (this one is added in new version of article in part in 2.12)
- every flow connection must relate to flow II $y_{i,s,j,i,s,j} = \sum_{u,v\in<1,n>} y_{u,p,v,i,s,j}$ for all $i,j\in<1,n>$ and $s\in<2,n>$ and $p\in<1,n-1>$ where $p<s$

After adding this two conditions instance (graph ABCD) from previous section was solved correctly.

Now let us go back to example of four valleys. After introduction of $y_{i,s,j,u,p,v}$ solution shown on Figure 4 is incorrect. But it can be made corrected after adding additional paths within valley B and C. It will be shown on two figures:

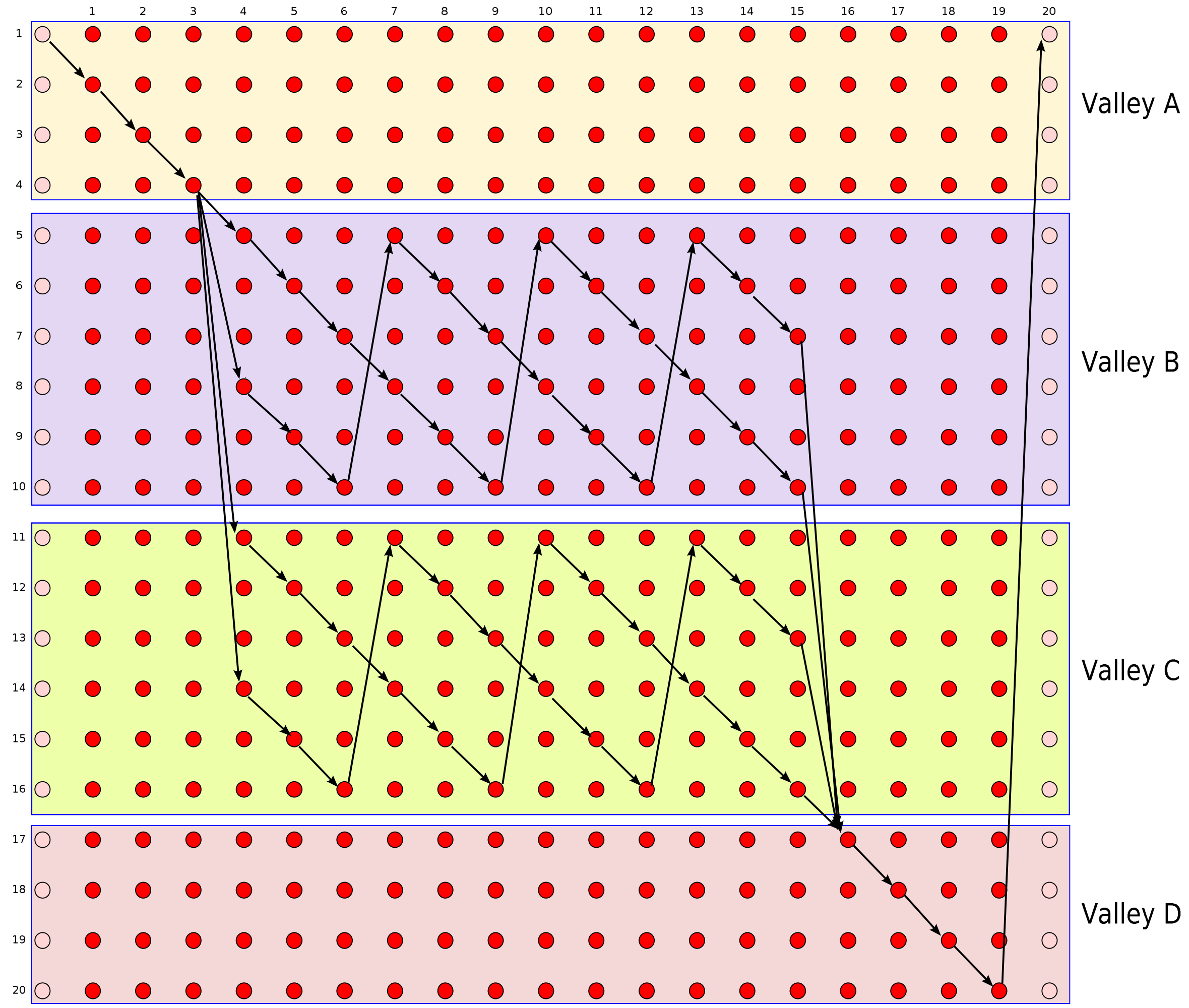

*Figure 5 Four valleys path on diagram – first paths*

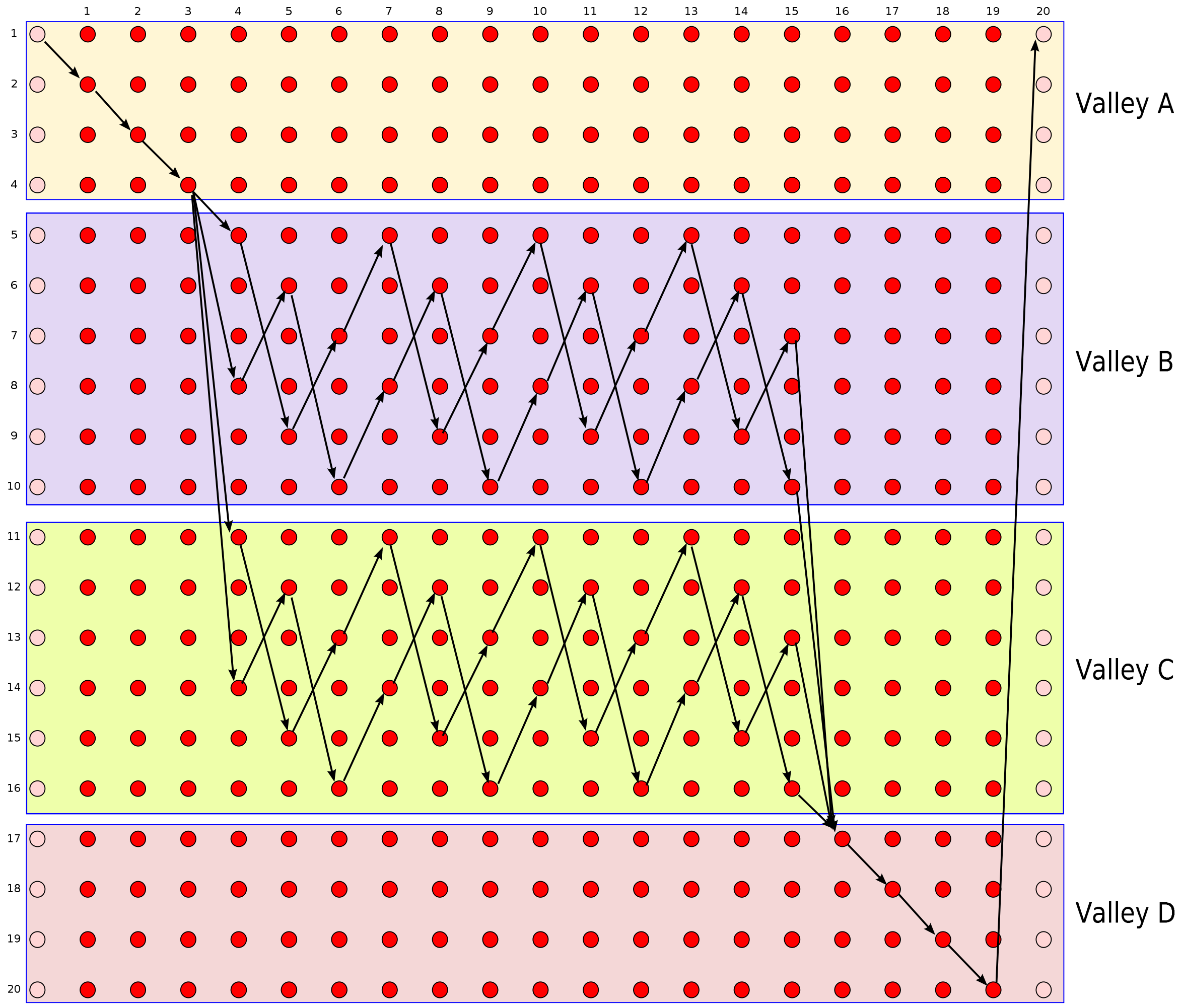

*Figure 6 Four valleys path on diagram – additional paths*

If we put this together we obtain solution similar to presented in previous section, but this time salesman is splitted to 8 parts and every town in valleys B and C are visited 8 times (on 4 stages 2/8 at one stage):

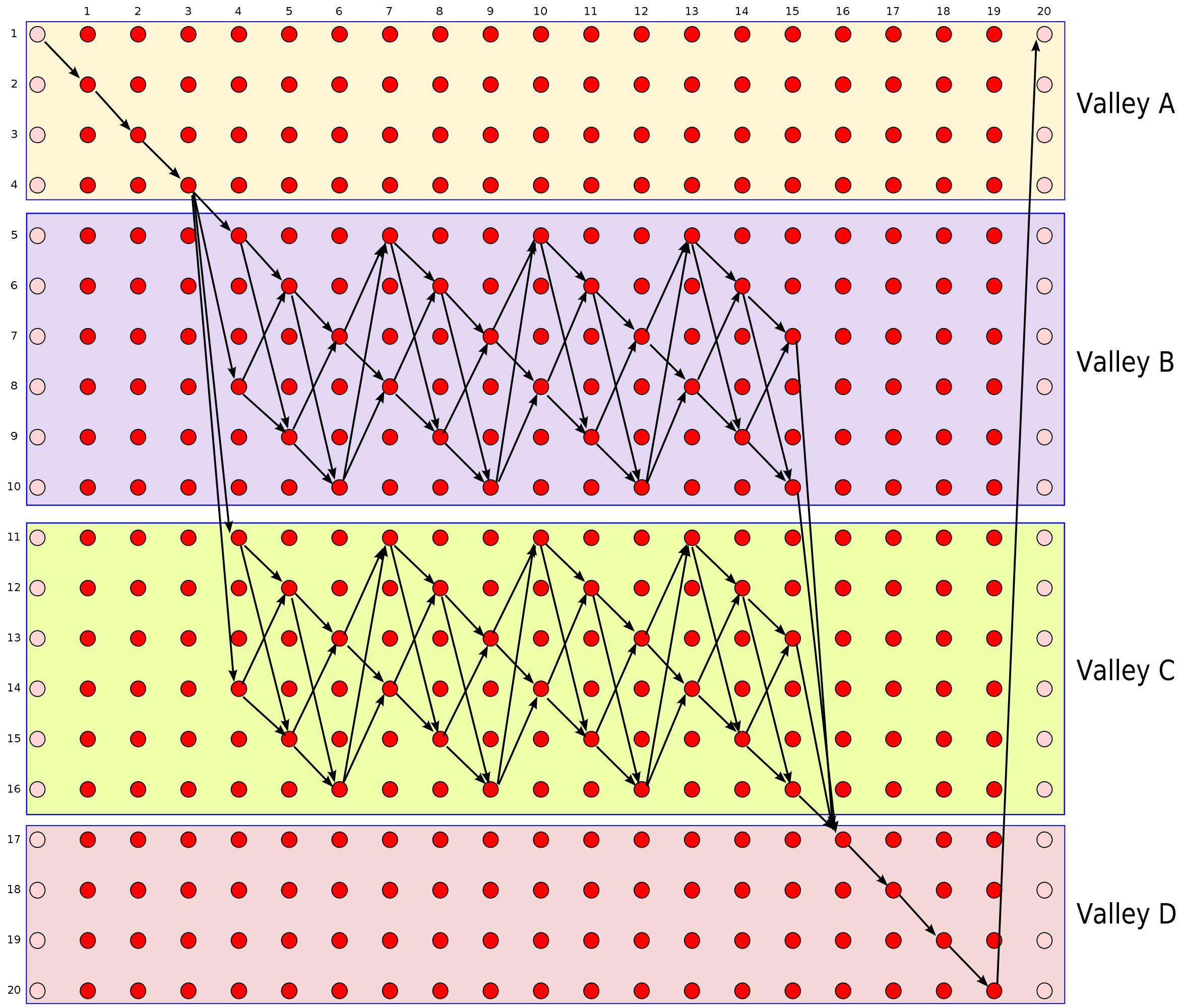

*Figure 7 All necessary paths for four valleys*

This solution is correct for all restrictions defined in article for $y_{i,s,j,u,p,v}$ but is incorrect solution for TSP. To understand better one should read explanation how to understand additional paths: they "switch" between two paths from valley in every step. If one denotes one path through valley B as B1 and second as B2 first of additional paths consist of nodes: $B1_1$, $B2_2$, $B1_3$, $B2_4$…

$x$ variables are set directly as shown on figures. In next step in order to build $y$ variables (with their values) algorithm should:

1. Pick up arc ($x_{i,s,j}$)
2. For every following stage ($p>s$)
   a) list every nodes in which flow may be at stage $p$-1
   b) list every nodes in which flow may be at stage $p$
   c) split flow from each node at stage $p$-1 to every node at stage $p$

We can see that because of two independent paths through valleys B and C together with additional switching between them for every arc from this valleys we can create consistent flow from entry to point where valley is leaved. For any arcs from valleys A and D their $y_{i,s,j,*,*,*}$ present whole graph.

One should understand why this solution is correct. We may change all restrictions mentioned in the beginning of this section into one sentence "every $x_{i,s,j}$ there must be followed by solution consisting of $x_{u,p,v}$ correct for $x$'s". For $x$'s model it was enough to have single path through valleys B and C (see figure 4) and now we have 4 different arcs at every stage in valleys B and C. It means that if we "pick up" any arc from one of this valleys we can have at least one path to exit from valley going through cities not visited in "picked" arc.

This solution can be "defeated" by additional restriction (corresponding to 2.25 from old version of article, 2.13 from new version): $\sum_{k\in<1,n>,r>s} y_{i,s,j,k,r,t} + \sum_{k\in<1,n>,r<s} y_{k,r,t,i,s,j} = y_{i,s,j,i,s,j}$ for each $i,j,t\in<1,n>$ and $s\in<1,n>$. This requirement forces to change statement from previous paragraph to: "for every $x_{i,s,j}$ there must be consistent solution of $x_{u,p,v}$ correct for $x$'s model with whole flow through $x_{i,s,j}$". To build proper counter example if all necessary requirements were defined we need to double example from figure 7:

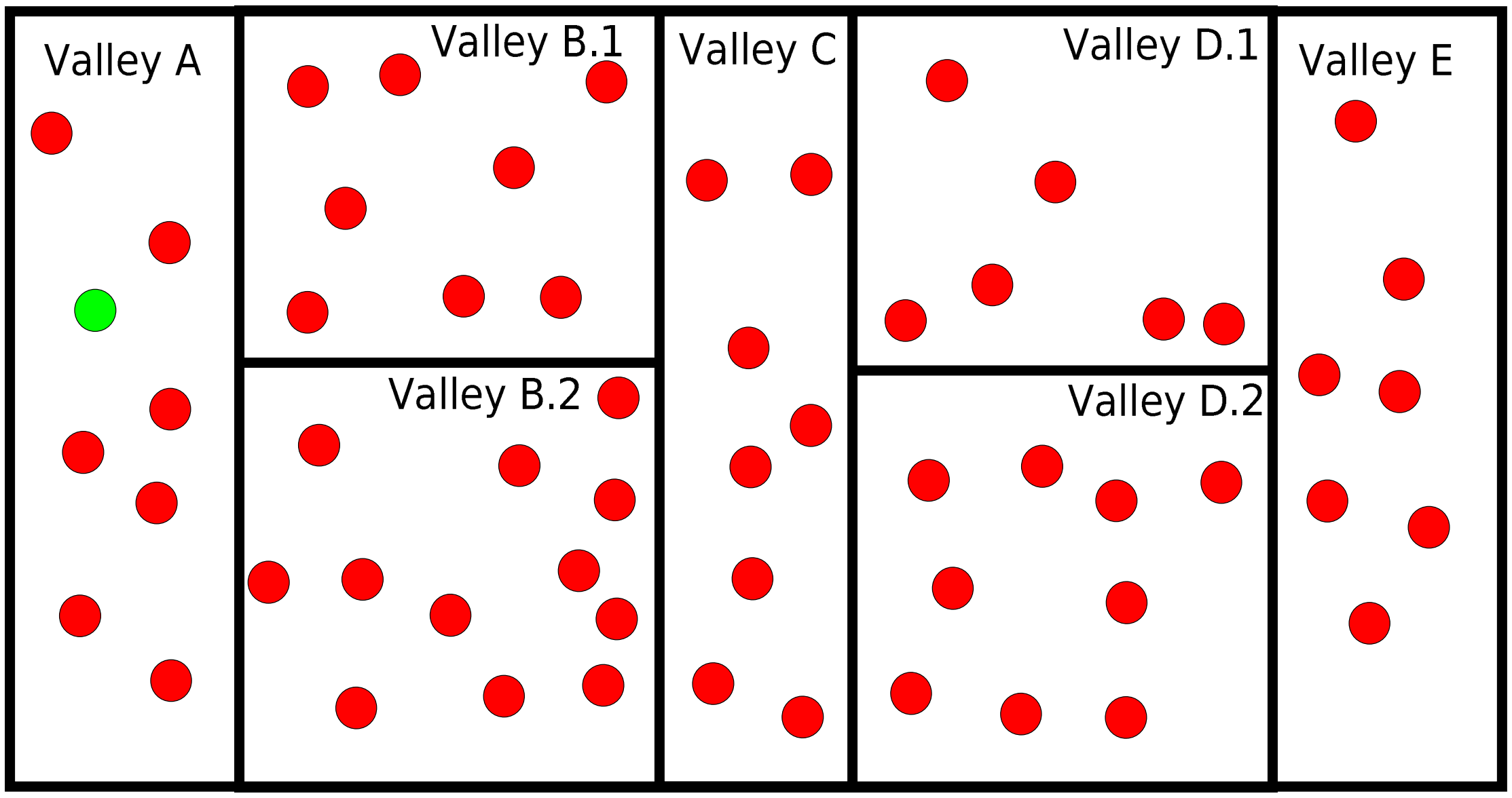

*Figure 8 Idea of seven valleys*

Now having in mind that for each arc we must have flow consistent for $x_{i,s,j}$ model we will build solution from 2 groups of paths (one should think about them in way showed on figure 7, but to show it in clear way some flows are omitted and only marked by color):

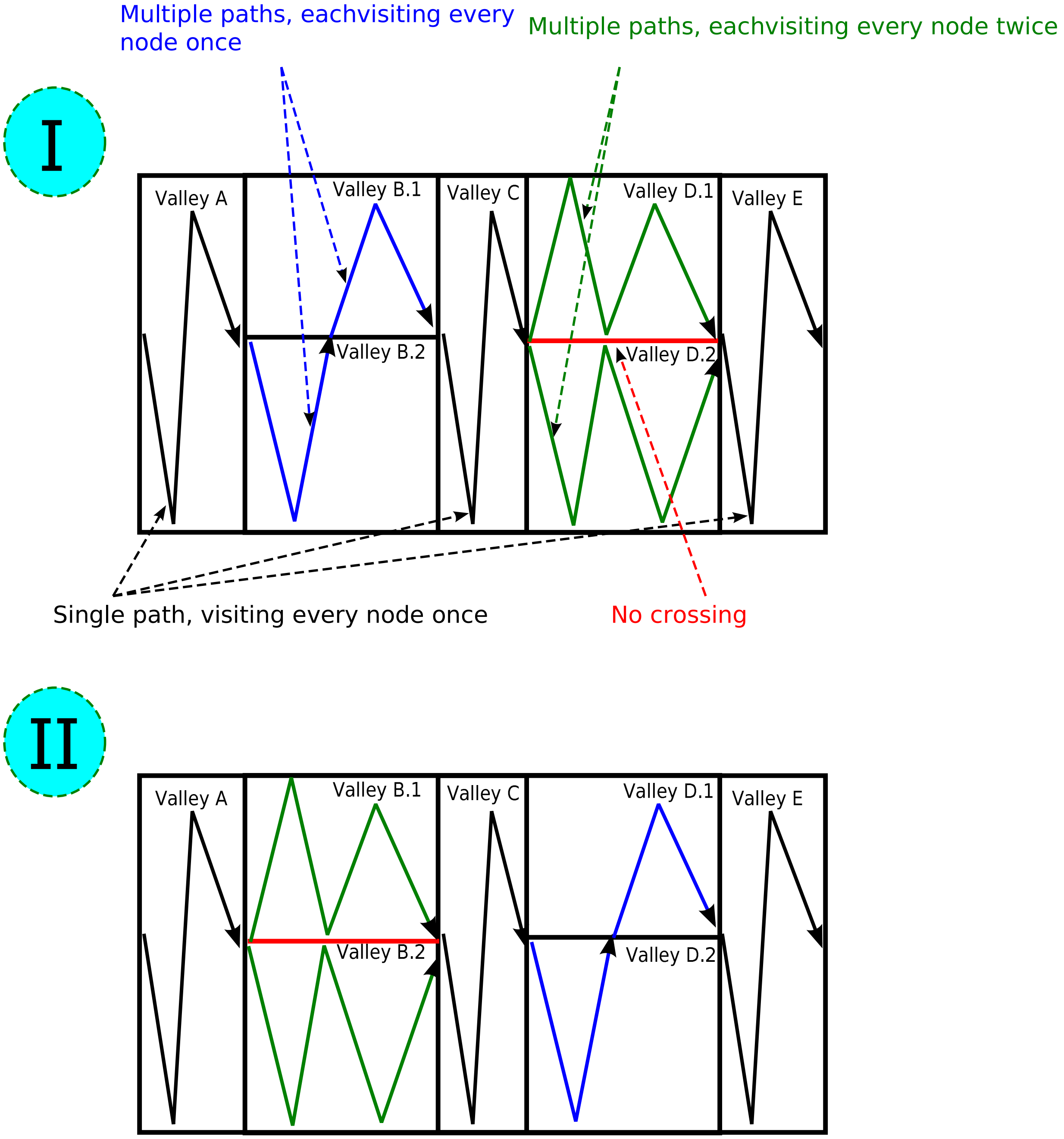

*Figure 9 Idea of paths for seven valleys*

We can see that on each picture we have correct flow through mountains in one pair of valleys and flow without crossing for second pair. If we built model for any $x_{i,s,j}$ then:

- if $x_{i,s,j}$ is from valley A, C or E then corresponding $x_{u,p,v}$ are equal to total flow for graph
- if $x_{i,s,j}$ is from valleys B.1 or B.2 (including arcs entering this valleys, crossing between them and leaving this valleys) then corresponding $x_{u,p,v}$ are equal to total flows presented on picture I
- if $x_{i,s,j}$ is from valleys D.1 or D.2 (including arcs entering this valleys, crossing between them and leaving this valleys) then corresponding $x_{u,p,v}$ are equal to total flows presented on picture II

Flow with no mountain crossing should be considered as presented on figure 7. Flow where mountains are crossed should be considered as on following figure:

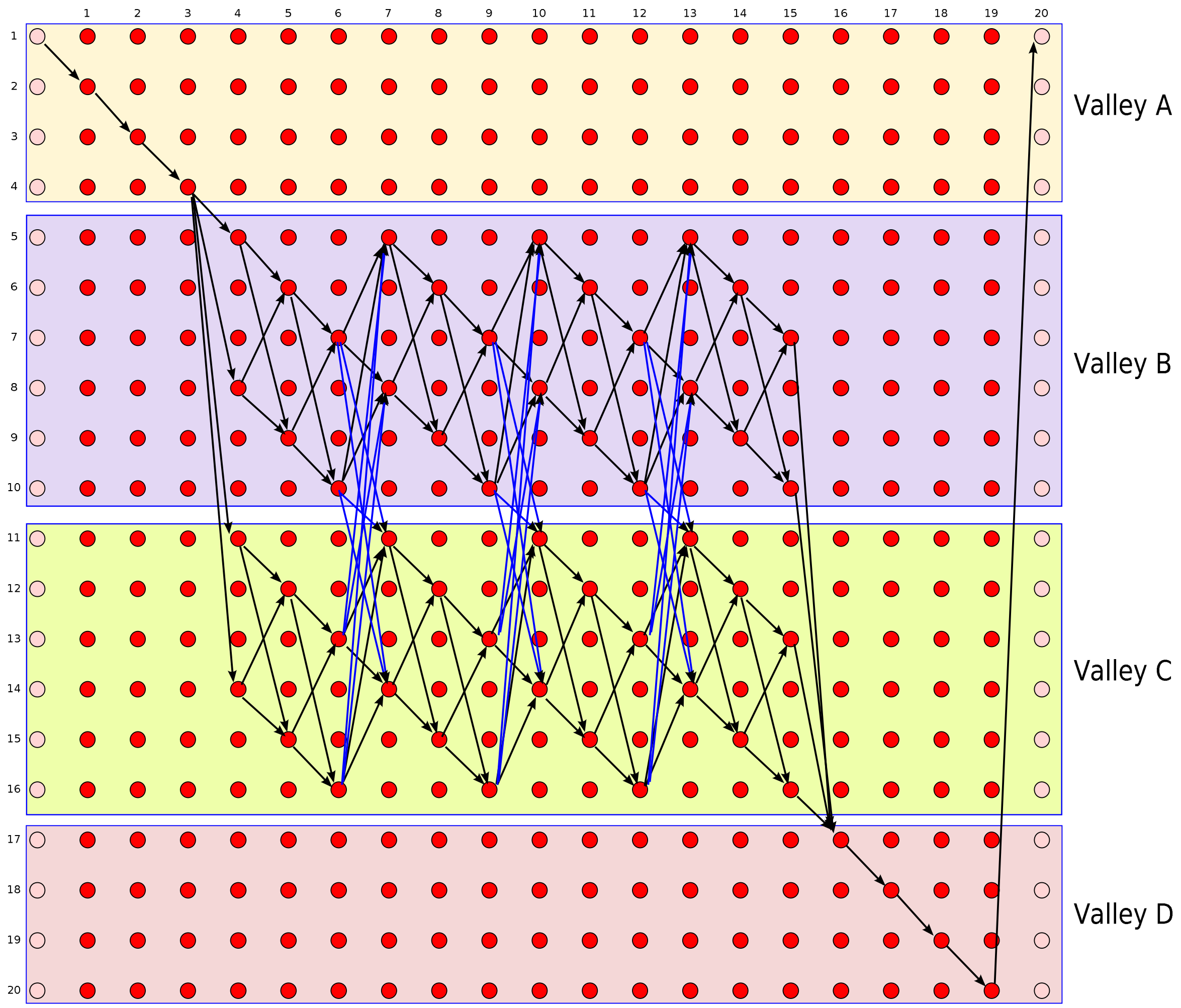

*Figure 10 Paths for four valleys with mountain crossing*

At every crossing 1/*x* of flow is considered to have crossed mountains, and total of crossings is equal to total flow.

Valleys B.1..D.2 with or without mountain crossing are similar having analogical internal flow and size. We can see that in such way we may make this 2-dimensional flow weaker – if we look from perspective of one dimension (by picking one arc) then in the same way model for one dimension was "bitten" we can have incorrect flow in other pair of valleys.

Is this incorrect solution of TSP problem? Of course – minimal flow for TSP problem would consist of 7 mountain crossings (A→B.1→B.2→C→D.1→D.2→E→A) and on every picture there has lower cost:

| | | |
|---|---|---|
| I | A→B.1→B.2→C | 3 crossings |
| | C→D.1 | ½ crossing |
| | C→D.2 | ½ crossing |
| | D.1→E | ½ crossing |
| | D.2→E | ½ crossing |
| | E→A | 1 crossing |
|  |  | Total: 6 crossings |
| II | A→B.1 | ½ crossing |
|  | A→B.2 | ½ crossing |
|  | B.1→C | ½ crossing |
|  | B.2→C | ½ crossing |
|  | C→D.1→D.2→E→A | 4 crossings |
|  |  | Total: 6 crossings |

If we combine this 2 groups of paths then overall cost is 6. This means that it is better then best solution for TSP. It has also possibility to respect every possible requirement defined for $y$ variables (not only those defined in article).

It is important that in new version of article this condition is removed, so presented in article model is even weaker and above considerations could be omitted, but are presented to show why such approach is incorrect in general.

## 2.3 $z_{i,s,j,u,p,v,k,r,t}$ variables (third dimension)

Author introduces variables $z_{i,s,j,u,p,v,k,r,t}$ which indexes can be interpreted as combination of three $x_{i,s,j}$ flow caused on $x_{k,r,t}$ by flow on $x_{u,p,v}$ caused originally by $x_{i,s,j}$ ($p>s$ and $r>p$). Again – to imagine this we may think of nesting table 1 in every cell of table 7 (it will not be presented here due to its size).

Having all restrictions from section 2.2 we have to add requirements for new variables:

- flow caused by two arcs have to be equal at every following stage: $\sum_{k,t\in<1,n>} z_{i,s,j,u,p,v,k,r,t} = y_{i,s,j,u,p,v}$ for each $i,j,u,v\in<1,n>$ and $s,p,r\in<1,n>$ having $s<p<r$ (equation 2.21 in old version of article, 2.9 in new version)
- flow caused by first to any arc must be equal to flow caused by first through other level arc: $\sum_{u,v\in<1,n>} z_{i,s,j,k,r,t,u,p,v} = y_{i,s,j,u,p,v}$ for each $i,j,k,t\in<1,n>$ and $s,p,r\in<1,n>$ having $s<p<r$ (equation 2.22 in article old version of article[7], 2.10 in new version)
- flow caused by first to any arc must be equal to flow caused by first through other level arc: $\sum_{i,j\in<1,n>} z_{i,s,j,k,r,t,u,p,v} = y_{u,p,v,k,r,t}$ for each $u,v,k,t\in<1,n>$ and $s,p,r\in<1,n>$ having $s<p<r$ (equation 2.23 in article old version of article[8], 2.11 in new version)
- flow caused on first level must be equal in following layers: $\sum_{k\in<1,n>,r\in<1,n>,r>s} z_{u,1,v,i,s,j,k,r,t} + \sum_{k\in<1,n>,r\in<1,n>,r<s} z_{u,1,v,t,r,k,i,s,j} = y_{u,1,v,i,s,j}$ for each $u,v,i,j,t\in<1,n>$ and $s\in<1,n>$ having $s>1$ (equation 2.25 in old version of article, 2.13 in new version)

Last equation could be even stronger if it was stated not only for $x_{u,1,v}$ but for any pair of arcs – but still it can be "defeated" what will be shown further on. In new version of article this condition is not implemented – author claims that "*Note that because of Proposition 4, the upper bounds on the $y_{irjkst}$ and $z_{irjupvkst}$ variables in constraints 2.18 are redundant in Problem BLP. Also, it is easy to observe that the visits requirements constraints 2.12 – 2.13 are not used (and hence, are not needed) in any of the proofs in section 2 of this paper. Hence, those constraints (i.e., constraints 2.12 - 2.13) are redundant in*

[7] In article there is error in his equation in old version of article – correct version was given on page 24
[8] In article there is error in his equation in old version of article – correct version was given on page 24

*Problem BLP (and therefore, in Problem IP). Hence, in implementing the model, we discarded constraints 2.12 – 2.13*". Skipping it makes counter example presented on figure 12 from this chapter be counter example for whole article!

Now getting back to example with four valleys. We said that model for $y$ was correct when for every $x_{i,s,j}$ there was possibility to prepare following (consistent) flow of $x_{u,p,v}$. We achieved that using two ways through valleys B and C and additional paths "switching" between this two paths at every stage. Building counter example for $z_{i,s,j,u,p,v,k,r,t}$ is more then similar. We may see that for every pair ($x_{i,s,j}$ and $x_{u,p,v}$ what is equivalent to $y_{i,s,j,u,p,v}$) we have to show consistent flow to the end. If we had three ways through valleys (with additional "switching" as well) we would be able to meet this requirement. Following pictures show this flow:

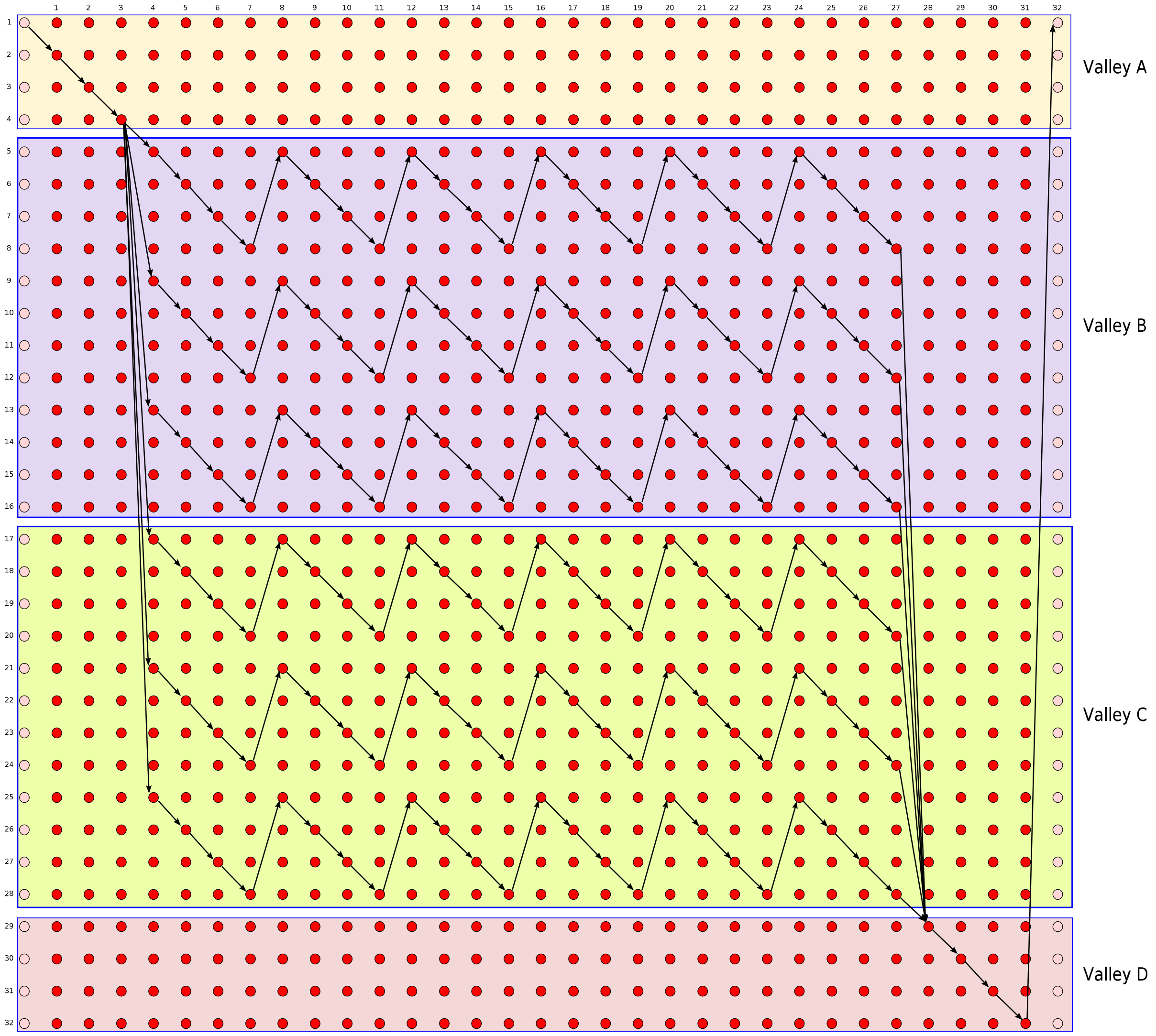

*Figure 11 Four valleys path on diagram – first paths*

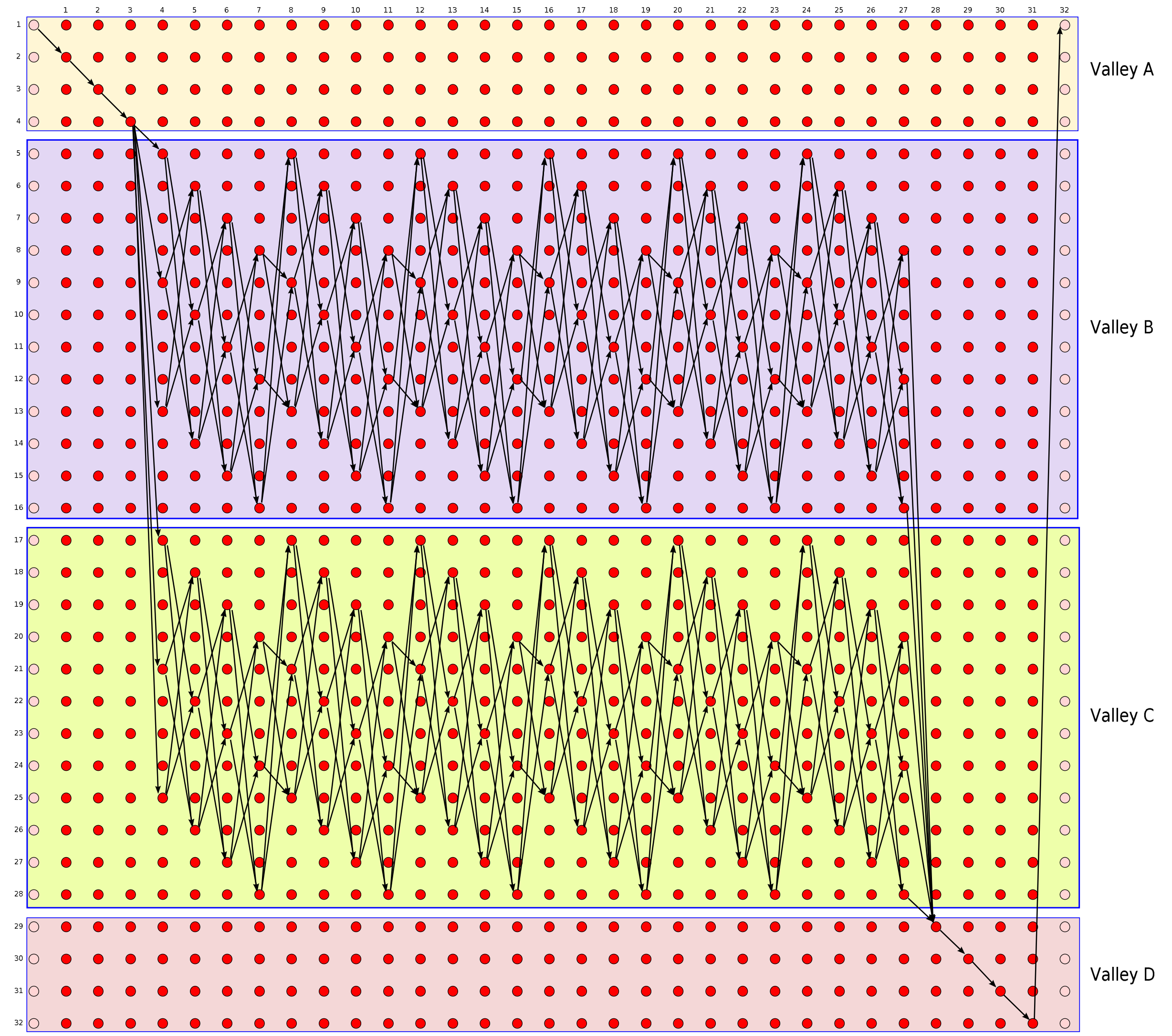

*Figure 12 Four valleys path on diagram – additional paths*

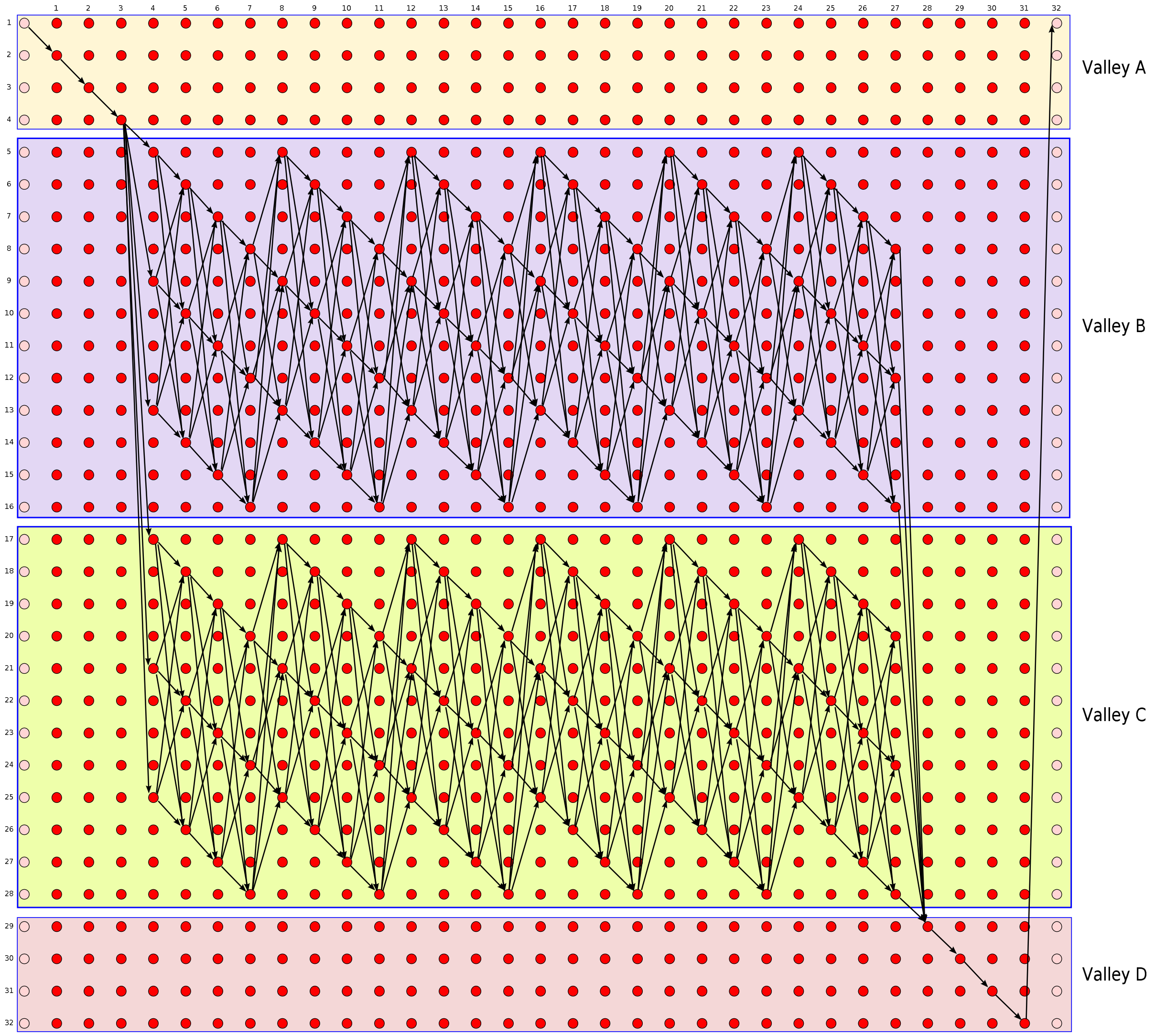

*Figure 13 All necessary paths for four valleys*

$x$ variables are set directly as shown on figures. In next step in order to build $y$ variables (with their values) algorithm should:

1. Pick up arc ($x_{i,s,j}$)
2. For every following stage ($p>s$)
   a) list every nodes in which flow from $x_{i,s,j}$ may be non-zero at stage $p$-1
   b) list every nodes in which flow from $x_{i,s,j}$ may be non-zero be at stage $p$
   c) split flow from each node at stage $p$-1 to every node at stage $p$

Then to build $z$ variables we do analogical thing:

1. Pick up arc ($x_{i,s,j}$)
2. For every following stage ($p>s$)
   a) Pick $x_{u,p,v}$ with non zero flow ($y_{i,s,j,u,p,v}$)
   b) For every following stage ($r>p$)
      i) list every nodes in which flow from $y_{i,s,j,u,p,v}$ may be non-zero at stage $r$-1
      ii) list every nodes in which flow from $y_{i,s,j,u,p,v}$ may be non-zero be at stage $r$
      iii) split flow from each node at stage $r$-1 to every node at stage $r$

We can see that because of three independent paths through valleys B and C together with additional switching between them for every pair of arcs from this valleys we can create consistent flow from entry to point where valley is leaved.

This algorithm is implemented in program listed in section 6.2.

As it can be seen instance is huge ($32^9$=35,184,372,088,832), but restricting variables only to arcs used (for every $y_{i,s,j,u,p,v}$ and $z_{i,s,j,u,p,v,k,r,t}$ we know that they are equal to 0 unless $x_{i,s,j}$, $x_{u,p,v}$ and $x_{k,r,t}$ are not equal to 0). List of non-zero variables with their values in text file for this instance takes "only" 33,647,527 bytes).

**After running checks for every condition from article defined in BLP model definition above instance meets all restrictions, so it is direct counter example for whole article!**

Of course if one defines every possible requirement (not only those from article) this instance could have been solved correctly. But for sure we can create counter example for all possible equations for $z$ variables. Method is analogical to this presented on figure 9, but this time we need to use three pairs of valleys, each containing 12 nodes.

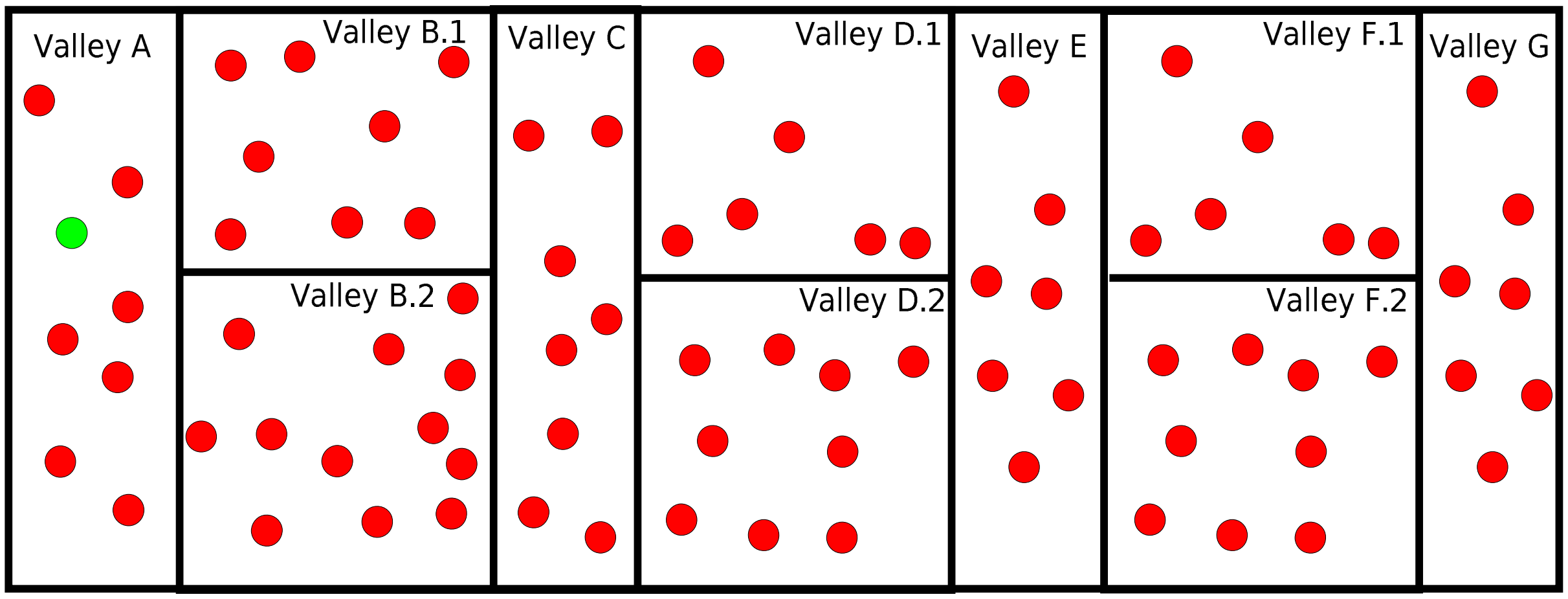

*Figure 14 Idea of ten valleys*

Now analogical to example from figure 9 we build total flow from 3 groups of paths. Each would contain 2 pairs with mountain crossings and one without crossing:

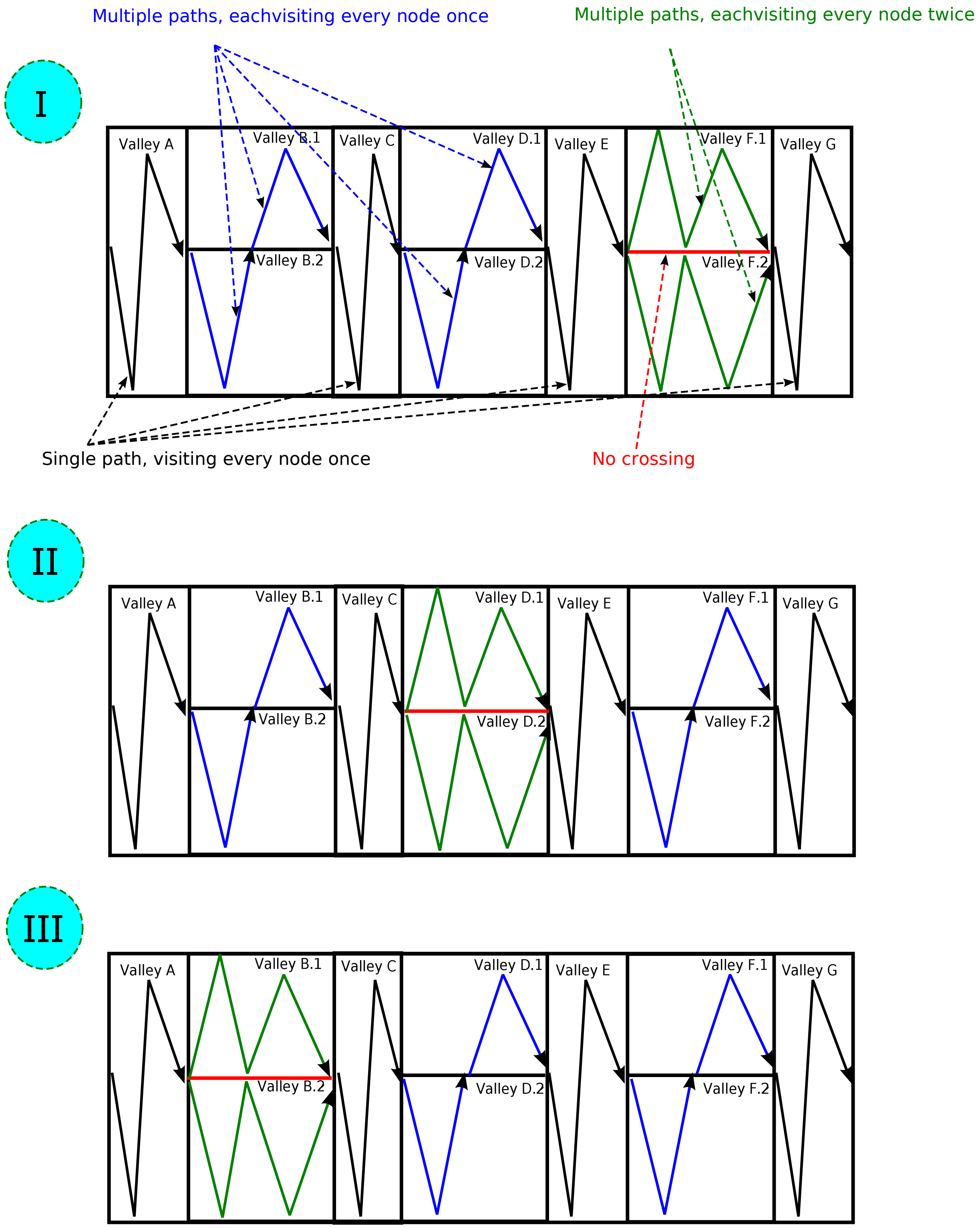

*Figure 15 Idea of ten valleys – three groups*

We can see that on each picture we have flow with mountain crossing in two pairs of valleys and without for third pair. If we build model for any $x_{i,s,j}$ and $x_{u,p,v}$ then:

- if $x_{i,s,j}$ and $x_{u,p,v}$ are both in valleys A, C, E or G then corresponding $x_{k,r,t}$ are equal to total flow for graph

- if one of $x_{i,s,j}$ or $x_{u,p,v}$ is in valleys A, C, E or G and second one is in any pair (B.X, D.X or F.X) then corresponding $x_{k,r,t}$ are equal to combination of flows which are correct for that pair (for example if $x_{i,s,j}$ is in valley A and $x_{u,p,v}$ is in D.1 then corresponding $x_{k,r,t}$ are combination of flow presented on pictures I and III)
- if $x_{i,s,j}$ and $x_{u,p,v}$ are both in the same pair of valleys (B.X, D.X or F.X) then corresponding $x_{k,r,t}$ are equal to combination of flows which are correct for that pair (for example if $x_{i,s,j}$ and $x_{u,p,v}$ are both in F.X then corresponding $x_{k,r,t}$ are combination of flow presented on pictures II and III)
- if one of $x_{i,s,j}$ or $x_{u,p,v}$ is in one pair of valleys (B.X, D.X or F.X) and second one is in different pair then corresponding $x_{k,r,t}$ are equal to combination of flows which are correct for these two pairs (for example if $x_{i,s,j}$ is in B.X and $x_{u,p,v}$ is in F.X then corresponding $x_{k,r,t}$ are combination of flow presented on picture II)

Valleys B.1..F.2 are similar to those presented on figure 13 having analogical internal flow and size, and crossings are analogical to those presented on figure 10 (there were 3 crossings, here will be 5). We can see that in such way we may make this 3-dimensional flow weaker – if we look from perspective of two dimensions (by picking two arcs) then in the same way model for one and two dimensions was "bitten" we can have incorrect flow in not chosen pair of valleys.

Is this incorrect solution for TSP problem? Of course – minimal flow for TSP problem would consist of 10 mountain crossings (A→B.1→B.2→C→D.1→D.2→E→F.1→F.2→G→A) and on every picture flow is of lower cost:

| | | |
|---|---|---|
| I | A→B.1→B.2→C→D.1→D.2→E | 6 crossings |
| | E→F.1 | ½ crossing |
| | E→F.2 | ½ crossing |
| | F.1→G | ½ crossing |
| | F.2→G | ½ crossing |
| | G→A | 1 crossing |
| | | Total: 9 crossings |
| II | A→B.1→B.2→C | 3 crossings |
| | C→D.1 | ½ crossing |
| | C→D.2 | ½ crossing |
| | D.1→E | ½ crossing |
| | D.2→E | ½ crossing |
| | E→F.1→F.2→G→A | 4 crossings |
| | | Total: 9 crossings |
| III | A→B.1 | ½ crossing |
| | A→B.2 | ½ crossing |
| | B.1→C | ½ crossing |
| | B.2→C | ½ crossing |
| | C→D.1→D.2→E→F.1→F.2→G→A | 7 crossings |
| | | Total: 9 crossings |

If we combine this 3 groups of paths then overall cost is 9. This means that it is better then best solution for TSP. It has also possibility to respect every possible requirement defined for $z$ variables (not only those defined in article).

Analogical counter example could be defined for any number of dimensions.

## 2.4 BLP problem definition

In section 3 (in old version 2.3 in new version) of original article there is a statement of BLP problem claimed to be equivalent to integer version of problem discussed above. Author strokes that implementations should be very careful but present some critical errors in statement of the problem which have to be corrected before implementation (this refers to old version of article):

- Restriction 3.3: for $s$=2 and $v$!=$i$ equality is incorrect – $s$ should start from index 3 (corrected in new version)
- Restriction 3.8: this restriction does not contain $j$ in its body so requiring $t$!=$j$ is incorrect (corrected in new version)
- Restriction 3.9: this restriction does not contain $j$ in its body so requiring $k$!=$j$ and $t$!=$j$ is incorrect (corrected in new version)
- Restriction 3.18: indexes for which equation is to be build contain $i$ and $j$ but in body this variables are not used – they should be changed to $k$ and $t$ which are used but undeclared (corrected in new version)
- Restriction 3.23: $s$ is said to be 1 to $n$-4 but because of variable $z_{u,1,v,i,s,j,k,r,t}$ with restriction that $v$!=$i$ minimum value of $s$ must be 3 (condition omitted in new version)

In tested implementation every flow was considered (also from node 1). In this case we have to allow variables $y_{i,1,*,*,n,1}$ and $z_{i,1,*,*,*,*,*,n,i}$, modify as they require that first variable differs from last and range of staging indexes (adding 1). This operation is fully legal on this model as it is equivalent to adding new node number 1.

Program in PERL language generating all necessary variables and every equation described in latest version of article is listed in section 6.2. It will generate 2 files – 33 MB of variables (1,120,335 of them), and 169 MB of equations (997,419 of them). Program checks every equation if it is correct, and of course every equation is. Running in debug mode can show checking of every equation and also difference between total cost of flow for counter example in comparison to minimal total cost of best TSP solution (of course counter example has lower overall cost).

## 3 Other way of looking on model

From the other hand we may think of presented in article model as about way of expressing some set of solutions at once. How many sets should be presented?

We know that there are $n!$ different Hamilton paths through graph (each can be considered as valid TSP tour). BLP model should then express any subset of such paths, as there may be 1 or $n!$ optimal solutions. This means that it should address 1 of $2^{n!}$ objects (cardinality of power set over $n!$). If we think about space required to store such definitions it is O[9]($\log(2^{n!})$)=O($n!$). This means that such representation is impossible to store on polynomially bounded space.

Model limited to constant number of variables cannot express such sets. For example in this case for 3 dimensional model we have $n^9$ variables (for $k$ dimensional model we would have $n^{3*k}$ variables). Number of different sets of TSP tours expressed by model is $X^{n^9}$ for some constant $X$. Even if $X \in <0, n!^c>$ for some constant $c$, then model cannot express every possible subset.

Proof:
Number of subsets from power set over $n!$ is greater then possible number of subsets presented by model:

$$2^{n!} > \left(n!^c\right)^{n^9} \Leftrightarrow 2^{n!} > n!^{c*n^9} \Leftrightarrow \log\left(2^{n!}\right) > \log\left(n!^{c*n^9}\right) \Leftrightarrow$$
$$n!*\log(2) > c*n^9*\log(n!) \Leftrightarrow n! > c*n^9*\log(n!)$$

Now we know that $n*\log(n)>\log(n!)$
if then:
$n!>c*n^9*n*\log(n) \Rightarrow n!>c*n^9*\log(n!)$

We will check then if $n!>c*n^9*n*\log(n)$
$n!>c*n^9*n*\log(n) \Leftrightarrow n!>c*n^{10}*\log(n) \Leftrightarrow ((n-10)!)*(n-9)*(n-8)....*n / n^{10} > c*\log(n)$

Now we take a look at this part: $(n-9)*(n-8)....*n / n^{10}$
$(n-9)*(n-8)....*n / n^{10} \Leftrightarrow (n^{10}+c_1*n^9+c_2*n^8....+c_9*n) / n^{10} \Leftrightarrow n^{10} / n^{10}+c_1*n^9 / n^{10}+....+c_9*n / n^{10} \Leftrightarrow$
$1+d_1+d_2....$

It is important that this expression is greater then 1. We know that:
$((n-10)!)*(n-9)*(n-8)....*n / n^{10} > (n-10)!$

So if $(n-10)! > c*\log(n)$ then: $n! / n^{10} > c*\log(n)$ and from above $2^{n!} > n!^{c*n^9}$ .

It is obvious that for $n$ growing to infinity for any constant $c$: $(n-10)! > c*\log(n)$, so whole considerations are true.

This means that for large $n$ model cannot express any subset of TSP tours – its expression capabilities are limited. If so, then for large $n$ it cannot prevent existence of feasible BLP solution consisting of some combination of paths neither of which is correct TSP solution, but together they make feasible solution.

Proposition 4 from new version of article is then false. In fact author does not prove that feasible solution for BLP cannot consist of paths being non-TSP tours.

---

[9] See http://en.wikipedia.org/wiki/Big_O_notation for definition of Big O

# 4 Summary

Looking back for optimal but incorrect solutions for $x_{i,s,j}$ and $y_{i,s,j,u,p,v}$ we can see that on $x_{i,s,j}$ we could build solutions with restrictions for one arc – we could require that sum of incoming flow to every node was the same for every node, and that at every stage in and out flows are equal. It was good enough if $x_{i,s,j}$ was 0 or 1 but after relaxing this assumption we saw solution for which every arc was correct, but their combination was incorrect. Model cannot restrict existence of such combination:

- $x_{1,1,2}$
- $x_{2,2,3}$
- $x_{3,3,2}$
- $x_{2,4,3}$
- $x_{3,5,2}$
- $x_{2,6,3}$
- …

Then we saw solution for $y_{i,s,j,u,p,v}$ – again after setting all necessary requirements we would obtain model with restrictions on each pair of arcs. This would be correct solution for $y_{i,s,j,u,p,v}$ equal to 0 or 1 but for decimal values we cannot prevent of existence of such variables:

- $y_{1,1,2,\mathbf{2,2,3}}$
- $y_{1,1,2,\mathbf{3,3,4}}$
- $y_{1,1,2,\mathbf{4,4,5}}$
- $y_{1,1,2,\mathbf{5,5,3}}$
- $y_{1,1,2,\mathbf{3,6,5}}$
- $y_{1,1,2,\mathbf{5,7,3}}$
- …

After "picking up" first arc model is no better then for $x_{i,s,j}$. Now building solution having all restrictions met for $x_{i,s,j}$ (equal to $y_{i,s,j,i,s,j}$) and $y_{i,s,j,u,p,v}$ is possible.

Adding next dimensions we have $z_{i,s,j,u,p,v,k,r,t}$ variables. Analogically to previous considerations we saw that this model restricts every triple of arcs. Solution for $z_{i,s,j,u,p,v,k,r,t}$ equal to 0 or 1 surely provides correct solution for original problem but for non-integer values we cannot prevent existence in solution of such combination:

- $z_{1,1,2,2,2,3,\mathbf{3,3,4}}$
- $z_{1,1,2,2,2,3,\mathbf{4,4,5}}$
- $z_{1,1,2,2,2,3,\mathbf{5,5,6}}$
- $z_{1,1,2,2,2,3,\mathbf{6,6,4}}$
- $z_{1,1,2,2,2,3,\mathbf{4,7,6}}$
- $z_{1,1,2,2,2,3,\mathbf{6,8,4}}$
- …

This means that after pointing two arcs remaining model is not stronger then for $x_{i,s,j}$. We also saw that such instance may have correct assignment with overall cost lower then optimal solution for TSP.

Adding next dimensions, picking up 4, 5 or even $k$ arcs and building restrictions between this variables cannot provide correct solver for TSP problem. Introducing next pairs of valleys enable to prepare counter example.

New version of article skips some of redundant equations (linearly dependant from other equations) and also skips very important condition 2.25 from old version (it is 2.13 in new version and reason why it is skipped is explained in the beginning of section 2.3 of this article). It allows whole solution to be proven incorrect with four valley example with three paths in each valley presented on figure 12 (source code in section 6.2). Getting this restriction back, or even making it stronger (we can require for every arc – not only first) to have corresponding pair with whole flow lead through this pair changes ability of model to correct instance size, but for *n* growing to infinity it cannot prevent finding incorrect solutions for integer version of problem being at same time correct for non-integer model.

This means that proposition 4 from new version of article is incorrect. It assumes that every feasible solution of BLP problem consists of valid TSP tours. This assumption is very important at every stage of proof of this proposition. What we have shown in this article is that feasible solution for BLP may consist of paths which in fact are not correct TSP paths.

Should this model be used anyway for small instances? In opinion of author of this paper not. If we use simple branch and bound method with some additional heuristics we can solve instances of size $n$=32 in couple of minutes, and it takes many hours only to generate equations for LP solver if we already know which variables are non-zero – if not then only generation of variables could last longer then $10^9$ years! So this method is much more expensive in terms of cost then simple B&B.

Exactly the same considerations apply to both articles from Moustapha Diaby [Dia1_2006], [Dia2_2006] and publicated recently article from Sergey Gubin [Gub2006] as they all use exactly same way of finding solution.

Why such approaches are incorrect? We can imagine matrix $n$ x $m$ size where we can define sum of each column and row – and question – do such equations suffice for finding value of each cell? In some cases of integer restricted solutions answer may be "yes", but for non-integer values we can have infinite number of assignments for variable values. This means that it is only matter of size when such approach can be proven being incorrect. All articles commented here use (maybe indirectly) such approach, and even if they are good enough for instances $n$=20, they are incorrect in general, what was proven by counter example given.

Dimensions of variables defines "look" of solution. If we refer only to $x_{i,s,j}$ then we can require that total flow "reaches" every city. This is enough for integer version of problem definition – every feasible solution is equivalent to valid TSP tour. After relaxing integer requirement this assumption (that feasible solution is equivalent to sum of TSP tours) is wrong.

After adding second dimension we obtain model in which we may require that flow through each arc belongs to solution correct in terms of first dimension (each city have to be "reached" with whole flow on this chosen arc). This is easy to beat if we have two pairs of valleys – in pair where arc was "picked up" we assume that solution is correct, but in other pair we can have exactly the same "trick" that was good enough to beat one dimensional model (in this model we cannot enforce flow between valleys from not chosen pair).

Finally after third dimension was added we have model requiring that total flow through any pair of arcs belongs to solution correct in terms of first dimension. This time we have to use three pairs of valleys because after "picking up" two arcs we can only enforce correct flow in up to two pairs, and third pair may contain exactly same "trick" which defeated one and two dimensional models (in this model we cannot enforce flow between valleys from not chosen pair).

Each level requires also more paths through valleys:

- for one dimension one path through each valley was enough
- for two dimensions we used two paths through valleys in pair (with switching between them – at every stage within valley we had $2*2^2$ arcs)
- for three dimensions we used three paths through valleys in pairs (with switching between them – at every stage within valley we had $2*3^3$ arcs)

This also require more cities in each valley: 3 for one dimension, 6 for two dimensions and 12 for three dimensions.

We can then see that every *k* dimensional model may be defeated by counter example containing:

- $1+(2+1)*k$ valleys
- each valley from pairs containing *k* paths
- each valley from pairs containing $3*2^{k-1}$ nodes

This means that counter example would have $(1+3*k)*3*2^{k-1}$ nodes.

But because in commented article not all possible requirements are stated we can see direct counter example with 32 nodes generated by program listed in section 6.2.

Resources for presented counter example may be downloaded from:
http://www.teycom.pl/diaby/diaby_counter_example.zip - program itself 5,7 kb
http://www.teycom.pl/diaby/variables_z.zip - variables with their values 7 Mb
http://www.teycom.pl/diaby/equations.zip – equations in LPT format 17 Mb

Referring to approach presented in section 4 we can see that model would be correct if it had strength to express any subset of valid TSP tours. But then it must have used:
- $2^{n!}$ Different values – it means that every value must have been stored on O(*n*!) bytes, or
- log(*n*) dimensions – this means that number of variables would be greater then $O(2^{n^{*c}})$

In both cases complexity of solution would not fit in definition of "polynomially bounded" so would not fit in P complexity class.

<u>Equality of **P** and **NP** complexity classes remain not proven.</u>

# 6 Appendix

## 6.1 List of $x_{i,s,j}$ for 4 valleys example with three paths

| Variable | Value | Variable | Value | Variable | Value | Variable | Value |
|---|---|---|---|---|---|---|---|
| x_1_1_2 | 18/18 | x_22_10_19 | 1/18 | x_24_16_17 | 1/18 | x_22_22_19 | 1/18 |
| x_2_2_3 | 18/18 | x_22_10_23 | 1/18 | x_24_16_21 | 1/18 | x_22_22_23 | 1/18 |
| x_3_3_4 | 18/18 | x_22_10_27 | 1/18 | x_24_16_25 | 1/18 | x_22_22_27 | 1/18 |
| x_4_4_13 | 3/18 | x_26_10_19 | 1/18 | x_28_16_17 | 1/18 | x_26_22_19 | 1/18 |
| x_4_4_17 | 3/18 | x_26_10_23 | 1/18 | x_28_16_21 | 1/18 | x_26_22_23 | 1/18 |
| x_4_4_21 | 3/18 | x_26_10_27 | 1/18 | x_28_16_25 | 1/18 | x_26_22_27 | 1/18 |
| x_4_4_25 | 3/18 | x_6_10_11 | 1/18 | x_8_16_13 | 1/18 | x_6_22_11 | 1/18 |
| x_4_4_5 | 3/18 | x_6_10_15 | 1/18 | x_8_16_5 | 1/18 | x_6_22_15 | 1/18 |
| x_4_4_9 | 3/18 | x_6_10_7 | 1/18 | x_8_16_9 | 1/18 | x_6_22_7 | 1/18 |
| x_13_5_10 | 1/18 | x_11_11_12 | 1/18 | x_13_17_10 | 1/18 | x_11_23_12 | 1/18 |
| x_13_5_14 | 1/18 | x_11_11_16 | 1/18 | x_13_17_14 | 1/18 | x_11_23_16 | 1/18 |
| x_13_5_6 | 1/18 | x_11_11_8 | 1/18 | x_13_17_6 | 1/18 | x_11_23_8 | 1/18 |
| x_17_5_18 | 1/18 | x_15_11_12 | 1/18 | x_17_17_18 | 1/18 | x_15_23_12 | 1/18 |
| x_17_5_22 | 1/18 | x_15_11_16 | 1/18 | x_17_17_22 | 1/18 | x_15_23_16 | 1/18 |
| x_17_5_26 | 1/18 | x_15_11_8 | 1/18 | x_17_17_26 | 1/18 | x_15_23_8 | 1/18 |
| x_21_5_18 | 1/18 | x_19_11_20 | 1/18 | x_21_17_18 | 1/18 | x_19_23_20 | 1/18 |
| x_21_5_22 | 1/18 | x_19_11_24 | 1/18 | x_21_17_22 | 1/18 | x_19_23_24 | 1/18 |
| x_21_5_26 | 1/18 | x_19_11_28 | 1/18 | x_21_17_26 | 1/18 | x_19_23_28 | 1/18 |
| x_25_5_18 | 1/18 | x_23_11_20 | 1/18 | x_25_17_18 | 1/18 | x_23_23_20 | 1/18 |
| x_25_5_22 | 1/18 | x_23_11_24 | 1/18 | x_25_17_22 | 1/18 | x_23_23_24 | 1/18 |
| x_25_5_26 | 1/18 | x_23_11_28 | 1/18 | x_25_17_26 | 1/18 | x_23_23_28 | 1/18 |
| x_5_5_10 | 1/18 | x_27_11_20 | 1/18 | x_5_17_10 | 1/18 | x_27_23_20 | 1/18 |
| x_5_5_14 | 1/18 | x_27_11_24 | 1/18 | x_5_17_14 | 1/18 | x_27_23_24 | 1/18 |
| x_5_5_6 | 1/18 | x_27_11_28 | 1/18 | x_5_17_6 | 1/18 | x_27_23_28 | 1/18 |
| x_9_5_10 | 1/18 | x_7_11_12 | 1/18 | x_9_17_10 | 1/18 | x_7_23_12 | 1/18 |
| x_9_5_14 | 1/18 | x_7_11_16 | 1/18 | x_9_17_14 | 1/18 | x_7_23_16 | 1/18 |
| x_9_5_6 | 1/18 | x_7_11_8 | 1/18 | x_9_17_6 | 1/18 | x_7_23_8 | 1/18 |
| x_10_6_11 | 1/18 | x_12_12_13 | 1/18 | x_10_18_11 | 1/18 | x_12_24_13 | 1/18 |
| x_10_6_15 | 1/18 | x_12_12_5 | 1/18 | x_10_18_15 | 1/18 | x_12_24_5 | 1/18 |
| x_10_6_7 | 1/18 | x_12_12_9 | 1/18 | x_10_18_7 | 1/18 | x_12_24_9 | 1/18 |
| x_14_6_11 | 1/18 | x_16_12_13 | 1/18 | x_14_18_11 | 1/18 | x_16_24_13 | 1/18 |
| x_14_6_15 | 1/18 | x_16_12_5 | 1/18 | x_14_18_15 | 1/18 | x_16_24_5 | 1/18 |
| x_14_6_7 | 1/18 | x_16_12_9 | 1/18 | x_14_18_7 | 1/18 | x_16_24_9 | 1/18 |
| x_18_6_19 | 1/18 | x_20_12_17 | 1/18 | x_18_18_19 | 1/18 | x_20_24_17 | 1/18 |
| x_18_6_23 | 1/18 | x_20_12_21 | 1/18 | x_18_18_23 | 1/18 | x_20_24_21 | 1/18 |
| x_18_6_27 | 1/18 | x_20_12_25 | 1/18 | x_18_18_27 | 1/18 | x_20_24_25 | 1/18 |
| x_22_6_19 | 1/18 | x_24_12_17 | 1/18 | x_22_18_19 | 1/18 | x_24_24_17 | 1/18 |
| x_22_6_23 | 1/18 | x_24_12_21 | 1/18 | x_22_18_23 | 1/18 | x_24_24_21 | 1/18 |
| x_22_6_27 | 1/18 | x_24_12_25 | 1/18 | x_22_18_27 | 1/18 | x_24_24_25 | 1/18 |
| x_26_6_19 | 1/18 | x_28_12_17 | 1/18 | x_26_18_19 | 1/18 | x_28_24_17 | 1/18 |
| x_26_6_23 | 1/18 | x_28_12_21 | 1/18 | x_26_18_23 | 1/18 | x_28_24_21 | 1/18 |
| x_26_6_27 | 1/18 | x_28_12_25 | 1/18 | x_26_18_27 | 1/18 | x_28_24_25 | 1/18 |
| x_6_6_11 | 1/18 | x_8_12_13 | 1/18 | x_6_18_11 | 1/18 | x_8_24_13 | 1/18 |
| x_6_6_15 | 1/18 | x_8_12_5 | 1/18 | x_6_18_15 | 1/18 | x_8_24_5 | 1/18 |
| x_6_6_7 | 1/18 | x_8_12_9 | 1/18 | x_6_18_7 | 1/18 | x_8_24_9 | 1/18 |
| x_11_7_12 | 1/18 | x_13_13_10 | 1/18 | x_11_19_12 | 1/18 | x_13_25_10 | 1/18 |
| x_11_7_16 | 1/18 | x_13_13_14 | 1/18 | x_11_19_16 | 1/18 | x_13_25_14 | 1/18 |
| x_11_7_8 | 1/18 | x_13_13_6 | 1/18 | x_11_19_8 | 1/18 | x_13_25_6 | 1/18 |
| x_15_7_12 | 1/18 | x_17_13_18 | 1/18 | x_15_19_12 | 1/18 | x_17_25_18 | 1/18 |
| x_15_7_16 | 1/18 | x_17_13_22 | 1/18 | x_15_19_16 | 1/18 | x_17_25_22 | 1/18 |
| x_15_7_8 | 1/18 | x_17_13_26 | 1/18 | x_15_19_8 | 1/18 | x_17_25_26 | 1/18 |
| x_19_7_20 | 1/18 | x_21_13_18 | 1/18 | x_19_19_20 | 1/18 | x_21_25_18 | 1/18 |
| x_19_7_24 | 1/18 | x_21_13_22 | 1/18 | x_19_19_24 | 1/18 | x_21_25_22 | 1/18 |
| x_19_7_28 | 1/18 | x_21_13_26 | 1/18 | x_19_19_28 | 1/18 | x_21_25_26 | 1/18 |
| x_23_7_20 | 1/18 | x_25_13_18 | 1/18 | x_23_19_20 | 1/18 | x_25_25_18 | 1/18 |
| x_23_7_24 | 1/18 | x_25_13_22 | 1/18 | x_23_19_24 | 1/18 | x_25_25_22 | 1/18 |
| x_23_7_28 | 1/18 | x_25_13_26 | 1/18 | x_23_19_28 | 1/18 | x_25_25_26 | 1/18 |
| x_27_7_20 | 1/18 | x_5_13_10 | 1/18 | x_27_19_20 | 1/18 | x_5_25_10 | 1/18 |
| x_27_7_24 | 1/18 | x_5_13_14 | 1/18 | x_27_19_24 | 1/18 | x_5_25_14 | 1/18 |
| x_27_7_28 | 1/18 | x_5_13_6 | 1/18 | x_27_19_28 | 1/18 | x_5_25_6 | 1/18 |
| x_7_7_12 | 1/18 | x_9_13_10 | 1/18 | x_7_19_12 | 1/18 | x_9_25_10 | 1/18 |
| x_7_7_16 | 1/18 | x_9_13_14 | 1/18 | x_7_19_16 | 1/18 | x_9_25_14 | 1/18 |
| x_7_7_8 | 1/18 | x_9_13_6 | 1/18 | x_7_19_8 | 1/18 | x_9_25_6 | 1/18 |
| x_12_8_13 | 1/18 | x_10_14_11 | 1/18 | x_12_20_13 | 1/18 | x_10_26_11 | 1/18 |
| x_12_8_5 | 1/18 | x_10_14_15 | 1/18 | x_12_20_5 | 1/18 | x_10_26_15 | 1/18 |
| x_12_8_9 | 1/18 | x_10_14_7 | 1/18 | x_12_20_9 | 1/18 | x_10_26_7 | 1/18 |

| Variable | Value | Variable | Value | Variable | Value | Variable | Value |
|---|---|---|---|---|---|---|---|
| x_16_8_13 | 1/18 | x_14_14_11 | 1/18 | x_16_20_13 | 1/18 | x_14_26_11 | 1/18 |
| x_16_8_5 | 1/18 | x_14_14_15 | 1/18 | x_16_20_5 | 1/18 | x_14_26_15 | 1/18 |
| x_16_8_9 | 1/18 | x_14_14_7 | 1/18 | x_16_20_9 | 1/18 | x_14_26_7 | 1/18 |
| x_20_8_17 | 1/18 | x_18_14_19 | 1/18 | x_20_20_17 | 1/18 | x_18_26_19 | 1/18 |
| x_20_8_21 | 1/18 | x_18_14_23 | 1/18 | x_20_20_21 | 1/18 | x_18_26_23 | 1/18 |
| x_20_8_25 | 1/18 | x_18_14_27 | 1/18 | x_20_20_25 | 1/18 | x_18_26_27 | 1/18 |
| x_24_8_17 | 1/18 | x_22_14_19 | 1/18 | x_24_20_17 | 1/18 | x_22_26_19 | 1/18 |
| x_24_8_21 | 1/18 | x_22_14_23 | 1/18 | x_24_20_21 | 1/18 | x_22_26_23 | 1/18 |
| x_24_8_25 | 1/18 | x_22_14_27 | 1/18 | x_24_20_25 | 1/18 | x_22_26_27 | 1/18 |
| x_28_8_17 | 1/18 | x_26_14_19 | 1/18 | x_28_20_17 | 1/18 | x_26_26_19 | 1/18 |
| x_28_8_21 | 1/18 | x_26_14_23 | 1/18 | x_28_20_21 | 1/18 | x_26_26_23 | 1/18 |
| x_28_8_25 | 1/18 | x_26_14_27 | 1/18 | x_28_20_25 | 1/18 | x_26_26_27 | 1/18 |
| x_8_8_13 | 1/18 | x_6_14_11 | 1/18 | x_8_20_13 | 1/18 | x_6_26_11 | 1/18 |
| x_8_8_5 | 1/18 | x_6_14_15 | 1/18 | x_8_20_5 | 1/18 | x_6_26_15 | 1/18 |
| x_8_8_9 | 1/18 | x_6_14_7 | 1/18 | x_8_20_9 | 1/18 | x_6_26_7 | 1/18 |
| x_13_9_10 | 1/18 | x_11_15_12 | 1/18 | x_13_21_10 | 1/18 | x_11_27_12 | 1/18 |
| x_13_9_14 | 1/18 | x_11_15_16 | 1/18 | x_13_21_14 | 1/18 | x_11_27_16 | 1/18 |
| x_13_9_6 | 1/18 | x_11_15_8 | 1/18 | x_13_21_6 | 1/18 | x_11_27_8 | 1/18 |
| x_17_9_18 | 1/18 | x_15_15_12 | 1/18 | x_17_21_18 | 1/18 | x_15_27_12 | 1/18 |
| x_17_9_22 | 1/18 | x_15_15_16 | 1/18 | x_17_21_22 | 1/18 | x_15_27_16 | 1/18 |
| x_17_9_26 | 1/18 | x_15_15_8 | 1/18 | x_17_21_26 | 1/18 | x_15_27_8 | 1/18 |
| x_21_9_18 | 1/18 | x_19_15_20 | 1/18 | x_21_21_18 | 1/18 | x_19_27_20 | 1/18 |
| x_21_9_22 | 1/18 | x_19_15_24 | 1/18 | x_21_21_22 | 1/18 | x_19_27_24 | 1/18 |
| x_21_9_26 | 1/18 | x_19_15_28 | 1/18 | x_21_21_26 | 1/18 | x_19_27_28 | 1/18 |
| x_25_9_18 | 1/18 | x_23_15_20 | 1/18 | x_25_21_18 | 1/18 | x_23_27_20 | 1/18 |
| x_25_9_22 | 1/18 | x_23_15_24 | 1/18 | x_25_21_22 | 1/18 | x_23_27_24 | 1/18 |
| x_25_9_26 | 1/18 | x_23_15_28 | 1/18 | x_25_21_26 | 1/18 | x_23_27_28 | 1/18 |
| x_5_9_10 | 1/18 | x_27_15_20 | 1/18 | x_5_21_10 | 1/18 | x_27_27_20 | 1/18 |
| x_5_9_14 | 1/18 | x_27_15_24 | 1/18 | x_5_21_14 | 1/18 | x_27_27_24 | 1/18 |
| x_5_9_6 | 1/18 | x_27_15_28 | 1/18 | x_5_21_6 | 1/18 | x_27_27_28 | 1/18 |
| x_9_9_10 | 1/18 | x_7_15_12 | 1/18 | x_9_21_10 | 1/18 | x_7_27_12 | 1/18 |
| x_9_9_14 | 1/18 | x_7_15_16 | 1/18 | x_9_21_14 | 1/18 | x_7_27_16 | 1/18 |
| x_9_9_6 | 1/18 | x_7_15_8 | 1/18 | x_9_21_6 | 1/18 | x_7_27_8 | 1/18 |
| x_10_10_11 | 1/18 | x_12_16_13 | 1/18 | x_10_22_11 | 1/18 | x_12_28_29 | 3/18 |
| x_10_10_15 | 1/18 | x_12_16_5 | 1/18 | x_10_22_15 | 1/18 | x_16_28_29 | 3/18 |
| x_10_10_7 | 1/18 | x_12_16_9 | 1/18 | x_10_22_7 | 1/18 | x_20_28_29 | 3/18 |
| x_14_10_11 | 1/18 | x_16_16_13 | 1/18 | x_14_22_11 | 1/18 | x_24_28_29 | 3/18 |
| x_14_10_15 | 1/18 | x_16_16_5 | 1/18 | x_14_22_15 | 1/18 | x_28_28_29 | 3/18 |
| x_14_10_7 | 1/18 | x_16_16_9 | 1/18 | x_14_22_7 | 1/18 | x_8_28_29 | 3/18 |
| x_18_10_19 | 1/18 | x_20_16_17 | 1/18 | x_18_22_19 | 1/18 | x_29_29_30 | 18/18 |
| x_18_10_23 | 1/18 | x_20_16_21 | 1/18 | x_18_22_23 | 1/18 | x_30_30_31 | 18/18 |
| x_18_10_27 | 1/18 | x_20_16_25 | 1/18 | x_18_22_27 | 1/18 | x_31_31_32 | 18/18 |
| | | | | | | x_32_32_1 | 18/18 |

*Table 8 List of $x_{i,s,j}$ variables with values for four valleys with three paths instance*

## 6.2 Generating all variables with values and equations for direct counter example for latest version of article

```
#* * * * * * * * * * * * * * * * * * * * * * * * * * * * * * * * * * * * * * */
#*                                                                            */
#*          This file is program generaing variables with values              */
#*    it produces 1 file with variables, and produces equations on STDOUT     */
#*    every equation can be evaluated in order to check if is fulfilled       */
#*    output file may be used as input file for SoPlex                        */
#*                                                                            */
#*    Copyright (C) 2006 Radoslaw Hofman                                      */
#*                        Poznan                                              */
#*    to run simply type at prompt                                            */
#*        perl LP_ce_soplex_x_double_new.pl > equations.lpt                   */
#*                                                                            */
#* * * * * * * * * * * * * * * * * * * * * * * * * * * * * * * * * * * * * * */
use strict;

my %variables;
my %numbers;
my %solution;
my %solution_x;
my $variables_no=1;
my $equations_no=0;

my $n=12;
```

```
my ($lead_in,$lead_out)=(4,4);
my $debug=0;
my $checking_eq=1;
my $checking_any=1;
my $restrict_boundaries=1;
my ($M,$R);
my $total_flow_constant=32*81;
$M=2*$n+$lead_in+$lead_out;
$R=2*$n+$lead_in+$lead_out;

#########################################################################################################
#########################################################################################################
#########################################################################################################
sub make_x_II
{
  for (my $s=1;$s<$lead_in;$s++)
  {
    print "stage $s\n" if ($debug);
    my $str1="x_$s\_$s\_".($s+1);
    $solution_x{$str1}+=1;
    print "$str1\n" if ($debug);
  }
  {
    print "stage $lead_in\n" if ($debug);
    my ($str1,$str2,$str3,$str4,$str5,$str6)=();
    $str1="x_".($lead_in)."_".($lead_in)."_".($lead_in+1);$solution_x{$str1}+=1;
    $str2="x_".($lead_in)."_".($lead_in)."_".($lead_in+$n/3+1);$solution_x{$str2}+=1;
    $str3="x_".($lead_in)."_".($lead_in)."_".($lead_in+2*$n/3+1);$solution_x{$str3}+=1;
    $str4="x_".($lead_in)."_".($lead_in)."_".($lead_in+$n+1);$solution_x{$str4}+=1;
    $str5="x_".($lead_in)."_".($lead_in)."_".($lead_in+4*$n/3+1);$solution_x{$str5}+=1;
    $str6="x_".($lead_in)."_".($lead_in)."_".($lead_in+5*$n/3+1);$solution_x{$str6}+=1;
    print "$str1\t$str2\t$str3\t$str4\t$str5\t$str6\n" if ($debug);
  }
  for (my $s=$lead_in+1;$s<=$lead_in+$n/3;$s++)
  {
    print "stage $s\n" if ($debug);
    my $step=1;
    $step=-$n/3+1 if ($s==$lead_in+$n/3);
    my ($str1,$str2,$str3,$str4,$str5,$str6,$str7,$str8,$str9)=();
    for (my $b=0;$b<=1;$b++)
    {
      my $bdd=$b*$n;
      for (my $a=0;$a<2*3-(($s==$lead_in+$n/3)?(1):(0));$a++)
      {
        my $add=$a*$n/3;
        $str1="x_".($s+$bdd)."_".($s+$add)."_".($s+$bdd+$step);$solution_x{$str1}+=1;
        $str2="x_".($s+$bdd)."_".($s+$add)."_".($s+$bdd+$n/3+$step);$solution_x{$str2}+=1;
        $str3="x_".($s+$bdd)."_".($s+$add)."_".($s+$bdd+2*$n/3+$step);$solution_x{$str3}+=1;
        $str4="x_".($s+$bdd+$n/3)."_".($s+$add)."_".($s+$bdd+$step);$solution_x{$str4}+=1;
        $str5="x_".($s+$bdd+$n/3)."_".($s+$add)."_".($s+$bdd+$n/3+$step);$solution_x{$str5}+=1;
        $str6="x_".($s+$bdd+$n/3)."_".($s+$add)."_".($s+$bdd+2*$n/3+$step);$solution_x{$str6}+=1;
        $str7="x_".($s+$bdd+2*$n/3)."_".($s+$add)."_".($s+$bdd+$step);$solution_x{$str7}+=1;
        $str8="x_".($s+$bdd+2*$n/3)."_".($s+$add)."_".($s+$bdd+$n/3+$step);$solution_x{$str8}+=1;
        $str9="x_".($s+$bdd+2*$n/3)."_".($s+$add)."_".($s+$bdd+2*$n/3+$step);$solution_x{$str9}+=1;
        print "$str1\t$str2\t$str3\t$str4\t$str5\t$str6\t$str7\t$str8\t$str9\n" if ($debug);
      }
    }
  }
  {
    print "stage ".($lead_in+2*$n)."\n" if ($debug);
    my ($str1,$str2,$str3,$str4,$str5,$str6)=();
    $str1="x_".($lead_in+$n/3)."_".($lead_in+2*$n)."_".($lead_in+2*$n+1);$solution_x{$str1}+=1;
    $str2="x_".($lead_in+2*$n/3)."_".($lead_in+2*$n)."_".($lead_in+2*$n+1);$solution_x{$str2}+=1;
    $str3="x_".($lead_in+$n)."_".($lead_in+2*$n)."_".($lead_in+2*$n+1);$solution_x{$str3}+=1;
    $str4="x_".($lead_in+4*$n/3)."_".($lead_in+2*$n)."_".($lead_in+2*$n+1);$solution_x{$str4}+=1;
    $str5="x_".($lead_in+5*$n/3)."_".($lead_in+2*$n)."_".($lead_in+2*$n+1);$solution_x{$str5}+=1;
    $str6="x_".($lead_in+2*$n)."_".($lead_in+2*$n)."_".($lead_in+2*$n+1);$solution_x{$str6}+=1;
    print "$str1\t$str2\t$str3\t$str4\t$str5\t$str6\n" if ($debug);
  }
  for (my $s=$lead_in+2*$n+1;$s<=$lead_in+2*$n+$lead_out;$s++)
  {
    print "stage $s\n" if ($debug);
    my $minus=0;
    $minus=$s if ($s==$lead_in+2*$n+$lead_out);
    my $str1="x_".($s)."_".($s)."_".($s-$minus+1);
    $solution_x{$str1}+=1;
    print "$str1\n" if ($debug);
```

```
  }
}

sub make_y_II
{
  my %levels;
  foreach (keys %solution_x)
  {
    my $key=$_;
    print ">>#>>$key>>$solution_x{$key}\n" if ($debug);
    if ($key=~/x\_(\d+)\_(\d+)\_(\d+)/)
    {
      my ($i,$s,$j)=($1,$2,$3);
      $levels{$s}{cnt}++;
      $levels{$s}{list}.=',' if ($levels{$s}{list});
      $levels{$s}{list}.=$key;
    }
  }
  #make yisjisj
  foreach (keys %solution_x)
  {
    my $key=$_;
    if ($key=~/x\_(\d+)\_(\d+)\_(\d+)/)
    {
      my ($i,$s,$j)=($1,$2,$3);
      $solution{"y\_$i\_$s\_$j\_$i\_$s\_$j"}=$total_flow_constant/$levels{$s}{cnt};
      print ">>y\_$i\_$s\_$j\_$i\_$s\_$j>>".($solution{"y\_$i\_$s\_$j\_$i\_$s\_$j"})."\n" if ($debug);
    }
  }
  #make yisjupv
  for (my $s=1;$s<$R;$s++)
  {
    for (my $p=$s+1;$p<=$R;$p++)
    {
      foreach (split(/\,/,$levels{$s}{list}))
      {
        my $src_arc=$_;
        my $cnt=0;
        my $list='';
        my ($ii,$ss,$jj)=($src_arc=~/x\_(\d+)\_(\d+)\_(\d+)/);
        print "checking for $src_arc list $levels{$p}{list}\n" if ($debug);
        foreach (split(/\,/,$levels{$p}{list}))
        {
          my $trg_arc=$_;
          my ($uu,$pp,$vv)=($trg_arc=~/x\_(\d+)\_(\d+)\_(\d+)/);
          print "checking $ii,$ss,$jj,$uu,$pp,$vv for src: $src_arc ($s/$p/$cnt)\n" if ($debug);
          if (check_y($ii,$ss,$jj,$uu,$pp,$vv))
          {
            $cnt++;
            $list.=',' if ($list);
            $list.=$trg_arc;
          }
        }
        die "No arcs found for $src_arc at level $p\n" unless ($cnt);
        foreach (split(/\,/,$list))
        {
          my $trg_arc=$_;
          my ($uu,$pp,$vv)=($trg_arc=~/x\_(\d+)\_(\d+)\_(\d+)/);
          $solution{"y\_$ii\_$ss\_$jj\_$uu\_$pp\_$vv"}=$solution{"y\_$ii\_$ss\_$jj\_$ii\_$ss\_$jj"}/$cnt;
          print
">>y\_$ii\_$ss\_$jj\_$uu\_$pp\_$vv>>".($solution{"y\_$ii\_$ss\_$jj\_$ii\_$ss\_$jj"})."/$cnt>>".($solution
{"y\_$ii\_$ss\_$jj\_$uu\_$pp\_$vv"})."\n" if ($debug);
        }
      }
    }
  }
}

sub make_z_II
{
  my %levels;
  foreach (keys %solution)
  {
    my $key=$_;
    if ($key=~/y\_(\d+)\_(\d+)\_(\d+)\_(\d+)\_(\d+)\_(\d+)/)
    {
      print ">>#>>$key>>$solution{$key}\n" if ($debug);
```

```
      my ($i,$s,$j,$u,$p,$v)=($1,$2,$3,$4,$5,$6);
      next if ($s==$p);
      $levels{$s}{$p}{cnt}++;
      $levels{$s}{$p}{list}.=',' if ($levels{$s}{$p}{list});
      $levels{$s}{$p}{list}.=$key;
    }
  }
  for (my $s=1;$s<$R-1;$s++)
  {
    for (my $p=$s+1;$p<$R;$p++)
    {
      for (my $r=$p+1;$r<=$R;$r++)
      {
        foreach (split(/\,/,$levels{$s}{$p}{list}))
        {
          my $src_y=$_;
          my $cnt=0;
          my $list='';
          my ($ii,$ss,$jj,$uus,$pps,$vvs)=($src_y=~/y\_(\d+)\_(\d+)\_(\d+)\_(\d+)\_(\d+)\_(\d+)/);
          print "checking for $src_y ($solution{$src_y}) list $levels{$p}{$r}{list}\n" if ($debug);
          foreach (split(/\,/,$levels{$p}{$r}{list}))
          {
            my $trg_y=$_;
            my ($uut,$ppt,$vvt,$kk,$rr,$tt)=($trg_y=~/y\_(\d+)\_(\d+)\_(\d+)\_(\d+)\_(\d+)\_(\d+)/);
            #print "TU: ($uut,$ppt,$vvt,$kk,$rr,$tt) (($uus==$uut)&&($pps==$ppt)&&($vvs==$vvt))\n" if
($debug);
            if (($uus==$uut)&&($pps==$ppt)&&($vvs==$vvt))
            {
              print "possible $src_y->$trg_y ($solution{$src_y})->($solution{$trg_y})\n" if ($debug);
              if (check_z($ii,$ss,$jj,$uus,$pps,$vvs,$kk,$rr,$tt))
              {
                print "OK\n" if ($debug);
                $cnt++;
                $list.=',' if ($list);
                $list.=$trg_y;
              }
            }
          }
          die "No Y found for $src_y at level $r\n" unless ($cnt);
          foreach (split(/\,/,$list))
          {
            my $trg_y=$_;
            my ($uut,$ppt,$vvt,$kk,$rr,$tt)=($trg_y=~/y\_(\d+)\_(\d+)\_(\d+)\_(\d+)\_(\d+)\_(\d+)/);

$solution{"z\_$ii\_$ss\_$jj\_$uus\_$pps\_$vvs\_$kk\_$rr\_$tt"}=$solution{"y\_$ii\_$ss\_$jj\_$uus\_$pps\_$
vvs"}/$cnt;
            print
">>z\_$ii\_$ss\_$jj\_$uus\_$pps\_$vvs\_$kk\_$rr\_$tt>>".($solution{"y\_$ii\_$ss\_$jj\_$uus\_$pps\_$vvs"})
."/$cnt>>".($solution{"z\_$ii\_$ss\_$jj\_$uus\_$pps\_$vvs\_$kk\_$rr\_$tt"})."\n" if ($debug);
          }
        }
      }
    }
  }
}

sub write_solution
{
  open(OUT,">$_[0]");
  foreach (keys %solution)
  {
    print OUT "$_\t$solution{$_}\n";
  }
  close(OUT);
}

sub read_solution
{
  open(IN,"<$_[0]");
  while(<IN>)
  {
    my $line=$_;
    $line=~s/[\r\n]//g;
    my ($variable,$value)=split(/\t/,$line);
    $solution{$variable}=$value;
    if ($variable=~/y\_(\d+)\_(\d+)\_(\d+)\_(\d+)\_(\d+)\_(\d+)/)
    {
```

```
      my ($i,$s,$j,$u,$p,$v)=($1,$2,$3,$4,$5,$6);
      $solution_x{"x_$i\_$s\_$j"}++;
      $solution_x{"x_$u\_$p\_$v"}++;
      die "Incorrect variable $variable\n" unless (check_y($i,$s,$j,$u,$p,$v));
    }
    elsif ($variable=~/z\_(\d+)\_(\d+)\_(\d+)\_(\d+)\_(\d+)\_(\d+)\_(\d+)\_(\d+)\_(\d+)/)
    {
      my ($i,$s,$j,$u,$p,$v,$k,$r,$t)=($1,$2,$3,$4,$5,$6,$7,$8,$9);
      $solution_x{"x_$i\_$s\_$j"}++;
      $solution_x{"x_$u\_$p\_$v"}++;
      $solution_x{"x_$k\_$r\_$t"}++;
      die "Incorrect variable $variable\n" unless (check_z($i,$s,$j,$u,$p,$v,$k,$r,$t));
    }
  }
  close(IN);
  print "Variables read\n" if ($debug);
}

my $file='variables_z.txt';

if ($checking_any)
{
  if (-e $file)
  {
    print "attept to read $file\n" if ($debug);
    read_solution($file);
  }
  else
  {
    print "NO FILE $file\n" if ($debug);
    make_x_II();
    make_y_II();
    make_z_II();
    write_solution($file);
  }
}

#########################################################################################################
#########################################################################################################
#########################################################################################################
sub get_variable_no
{
  unless ($variables{$_[0]})
  {
    $variables{$_[0]}=$variables_no;
    $numbers{$variables_no}=$_[0];
    $variables_no++;
  }
  return $variables{$_[0]};
}

sub check_z
{
  my ($i,$s,$j,$u,$p,$v,$k,$r,$t)=@_;
  return 0 if (($s>=$p)||($p>=$r)||($s>=$r));
  return 0 unless check_y($i,$s,$j,$u,$p,$v);
  return 0 unless check_y($i,$s,$j,$k,$r,$t);
  return 0 unless check_y($u,$p,$v,$k,$r,$t);
  return 1;
}

sub check_y
{
  my ($i,$s,$j,$u,$p,$v)=@_;
  return 0 unless check_x($i,$s,$j);
  return 0 unless check_x($u,$p,$v);
  #return 0 unless check_any_y($i,$s,$j,$u,$p,$v);
  #2.26 part 1
  return 0 if ($p<$s);
  #2.26 part 2
  return 0 if (($p==$s)&&(($i!=$u)||($j!=$v)));
  #2.26 part 3
  return 0 if (($p==$s+1)&&($j!=$u));
  #2.26 part 4
  return 0 if (($p>$s)&&(($s>1)||($p<$R))&&($i==$v));
  #2.26 part 5
  return 0 if (($p>$s)&&($i==$u));
```

```
  #2.26 part 6
  return 0 if (($p>$s)&&($j==$v));
  #2.26 part 7
  return 0 if (($p>$s+1)&&($j==$u));
  #2.26 part 8
  return 0 if ($i==$j);
  #2.26 part 9
  return 0 if ($u==$v);
  if (($s>$lead_in)&&($s<$p)&&($p<=$lead_in+2*$n)&&(
    (($i<=$lead_in+$n)&&($u>$lead_in+$n))
    ||
    (($u<=$lead_in+$n)&&($i>$lead_in+$n))
  ))
  {
    print "Incorrect A ($i,$s,$j,$u,$p,$v)\n" if ($debug);
    return 0;
  }
  if (($s==$lead_in)&&($s<$p)&&($p<=$lead_in+2*$n)&&(
    (($j<=$lead_in+$n)&&($u>$lead_in+$n))
    ||
    (($u<=$lead_in+$n)&&($j>$lead_in+$n))
  ))
  {
    print "Incorrect B ($i,$s,$j,$u,$p,$v)\n" if ($debug);
    return 0;
  }
  return 1;
}

sub check_x
{
  my ($i,$s,$j)=@_;
  return 0 if (($i>$M)||($s>$R)||($j>$M));
  return 0 if (($i<1)||($s<1)||($j<1));
  return 0 if ($i==$j);
  return 0 unless (check_any_x($i,$s,$j));
  return 1;
}

sub check_any_y
{
  my ($i,$s,$j,$u,$p,$v)=@_;
  return 0 unless (check_any_x($i,$s,$j));
  return 0 unless (check_any_x($u,$p,$v));
  return 0 unless ($solution{"y_$i\_$s\_$j\_$u\_$p\_$v"});
  return 1;
}

sub check_any_x
{
  my ($i,$s,$j)=@_;
  return 1 unless ($checking_any);
  my ($digit_cnt,$asterisk_cnt)=(0,0);
  if ($i ne '*')
  {
    if (($i<1)||($i>$M))
    {
      return 0;
    }
    else
    {
      $digit_cnt++;
    }
  }
  else
  {
    $asterisk_cnt++;
  }
  if ($s ne '*')
  {
    if (($s<1)||($s>$R))
    {
      return 0;
    }
    else
    {
      $digit_cnt++;
```

```
    }
  }
  else
  {
    $asterisk_cnt++;
  }
  if ($j ne '*')
  {
    if (($j<1)||($j>$M))
    {
      return 0;
    }
    else
    {
      $digit_cnt++;
    }
  }
  else
  {
    $asterisk_cnt++;
  }
  if ($digit_cnt==3)
  {
    return 0 unless ($solution_x{"x_$i\_$s\_$j"});
    return 1;
  }
  if ($asterisk_cnt>=2)
  {
    return 1;
  }
  my $ser_str='^x\_';
  if ($i>0)
  {
    $ser_str.=$i.'\_';
  }
  else
  {
    $ser_str.='\d+\_';
  }
  if ($s>0)
  {
    $ser_str.=$s.'\_';
  }
  else
  {
    $ser_str.='\d+\_';
  }
  if ($j>0)
  {
    $ser_str.=$j.'$';
  }
  else
  {
    $ser_str.='\d+$';
  }
  foreach (keys %solution_x)
  {
    my $key=$_;
    if ($key=~/$ser_str/)
    {
      return 1;
    }
  }
  return 0;
}

sub add_variable
{
  my ($sign,$factor,$variable,@params)=@_;
  $factor='' if ($factor==1);
  $factor.=' ' if ($factor);
  if (($variable eq 'y')&&(check_y(@params)))
  {
    my $str="y_$params[0]\_$params[1]\_$params[2]\_$params[3]\_$params[4]\_$params[5]";
    $variables{$str}++;
    return " $sign$factor$str";
  }
```

```
  elsif (($variable eq 'z')&&(check_z(@params)))
  {
    my
$str="z_$params[0]\_$params[1]\_$params[2]\_$params[3]\_$params[4]\_$params[5]\_$params[6]\_$params[7]\_$
params[8]";
    $variables{$str}++;
    return " $sign$factor$str";
  }
  return '';
}

sub check_equation
{
  return 1 unless ($checking_eq);
  print ">>CHKEQ - checking $_[0]\n" if ($debug);
  my $equation=$_[0];
  my $dont_die=$_[1];
  $equation=~s/\n/ /g;
  my $sign='eq';
  $sign='le' if ($equation=~/\<\=/);
  $sign='ge' if ($equation=~/\>\=/);
  my ($lewa,$prawa)=split(/[\=\<\>]+/,$equation);
  my $sum=0;
  while ($lewa=~/([\+\-])?\s*([\d\.]+)?\s*([yz][\d\_]+)/g)
  {
    my ($znak,$factor,$zmienna)=($1,$2,$3);
    $znak='+' unless ($znak);
    $factor=1 unless ($factor);
    my $local=$factor*$solution{$zmienna};
    if ($local)
    {
      if ($znak eq '-')
      {
        $sum-=$local;
      }
      else
      {
        $sum+=$local;
      }
    }
    print ">>$znak,$factor,$zmienna (solution:$solution{$zmienna},local:$local,acc:$sum)\n" if ($debug);
  }
  my $ok=0;
  if ($sign eq 'eq')
  {
    print "Checking equality $sum==$prawa\n" if ($debug);
    if (int($sum)==int($prawa))
    {
      $ok=1;
    }
  }
  elsif ($sign eq 'le')
  {
    print "Checking le equality $sum<=$prawa\n" if ($debug);
    if (int($sum)<=int($prawa))
    {
      $ok=1;
    }
  }
  elsif ($sign eq 'ge')
  {
    print "Checking le equality $sum>=$prawa\n" if ($debug);
    if (int($sum)>=int($prawa))
    {
      $ok=1;
    }
  }
  print "Equality violated!!\n" if (($debug)&&(!$ok));
  die "Equality violated!!\n" if ((!$ok)&&(!$dont_die));
  return $ok;
}

##########################################################################################
##########################################################################################
##########################################################################################
sub get_cost
{
```

```
  my ($i,$j)=@_;
  my ($i_valley,$j_valley)=(0,0);
  $i_valley='A' if (($i>=1)&&($i<=$lead_in));
  $i_valley='B' if (($i>=$lead_in+1)&&($i<=$lead_in+$n));
  $i_valley='C' if (($i>=$lead_in+$n+1)&&($i<=$lead_in+2*$n));
  $i_valley='D' if (($i>=$lead_in+2*$n+1)&&($i<=$lead_in+2*$n+$lead_out));
  $j_valley='A' if (($j>=1)&&($j<=$lead_in));
  $j_valley='B' if (($j>=$lead_in+1)&&($j<=$lead_in+$n));
  $j_valley='C' if (($j>=$lead_in+$n+1)&&($j<=$lead_in+2*$n));
  $j_valley='D' if (($j>=$lead_in+2*$n+1)&&($j<=$lead_in+2*$n+$lead_out));
  return 1 if ($i_valley eq $j_valley);
  return 1000;
}

goto tu if ($debug);

##########################################################################################################
#target cost
tu:
print "Minimize cost: \n";
my $string='';
my $last_cut=0;
for (my $i=1;$i<=$M;$i++)
{
  next unless (check_any_x($i,'*','*'));
  for (my $r=1;$r<=$R;$r++)
  {
    next unless (check_any_x($i,$r,'*'));
    for (my $j=1;$j<=$M;$j++)
    {
      next unless (check_any_x($i,$r,$j));
      $string.=add_variable('+',get_cost($i,$j),'y',$i,$r,$j,$i,$r,$j);
      $last_cut=length($string), $string.="\n"  if (length($string)-$last_cut>200);
    }
  }
}
$string=~s/^\s*\+//;
print "$string\n";
$equations_no++;
if ($debug)
{
  check_equation($string.">=".($total_flow_constant*4*1000),1);
}

print "Subject to\n";

#3.2 in new version
$string='';
$last_cut=0;
for (my $i=1;$i<=$M;$i++)
{
  next unless (check_any_x($i,1,'*'));
  for (my $j=1;$j<=$M;$j++)
  {
    next if ($i==$j);
    next unless (check_any_x($i,1,$j));
    $string.=add_variable('+',1,'y',$i,1,$j,$i,1,$j);
    $last_cut=length($string), $string.="\n"  if (length($string)-$last_cut>200);
  }
}
$string=~s/^\s*\+//;
$string.="=$total_flow_constant";
print "R302_$equations_no: $string\n";
check_equation($string);
$equations_no++;

#3.3 in new version
for (my $i=1;$i<=$M;$i++)
{
  next unless (check_any_x($i,2,'*'));
  for (my $j=1;$j<=$M;$j++)
  {
    next if ($i==$j);
    next unless (check_any_x($i,2,$j));
    $string=add_variable('+',1,'y',$i,2,$j,$i,2,$j);
    $last_cut=0;
    for (my $u=1;$u<=$M;$u++)
```

```
    {
      next if (($u==$i)||($u==$j));
      next unless (check_any_x($u,1,$i));
      $string.=add_variable('-',1,'y',$u,1,$i,$i,2,$j);
      $last_cut=length($string), $string.="\n"  if (length($string)-$last_cut>200);
    }
    $string=~s/^\s*\+//;
    $string.="=0";
    print "R303_$equations_no: $string\n";
    check_equation($string);
    $equations_no++;
  }
}

#3.4 in new version
for (my $i=1;$i<=$M;$i++)
{
  next unless (check_any_x($i,'*','*'));
  for (my $j=1;$j<=$M;$j++)
  {
    next if ($i==$j);
    next unless (check_any_x($i,'*',$j));
    for (my $r=3;$r<=$R;$r++)
    {
      next unless (check_any_x($i,$r,$j));
      $string=add_variable('+',1,'y',$i,$r,$j,$i,$r,$j);
      $last_cut=0;
      for (my $u=1;$u<=$M;$u++)
      {
        next if (($u==$i)||(($u==$j)&&($r<$R)));
        next unless (check_any_x($u,1,'*'));
        for (my $v=1;$v<=$M;$v++)
        {
          next if (($v==$i)||($v==$j)||($v==$u));
          next unless (check_any_x($u,1,$v));
          $string.=add_variable('-',1,'y',$u,1,$v,$i,$r,$j);
          $last_cut=length($string), $string.="\n"  if (length($string)-$last_cut>200);
        }
      }
      $string=~s/^\s*\+//;
      $string.="=0";
      print "R304_$equations_no: $string\n";
      check_equation($string);
      $equations_no++;
    }
  }
}

#3.5 in new version
for (my $i=1;$i<=$M;$i++)
{
  next unless (check_any_x($i,'*','*'));
  for (my $j=1;$j<=$M;$j++)
  {
    next unless (check_any_x($i,'*',$j));
    next if ($i==$j);
    for (my $r=1;$r<=$R-3+1;$r++)
    {
      next unless (check_any_x($i,$r,$j));
      $string=add_variable('+',1,'y',$i,$r,$j,$i,$r,$j);
      $last_cut=0;
      for (my $t=1;$t<=$M;$t++)
      {
        next unless (check_any_x($j,$r+1,$t));
        next if (($t==$i)||($t==$j));
        $string.=add_variable('-',1,'y',$i,$r,$j,$j,$r+1,$t);
        $last_cut=length($string), $string.="\n"  if (length($string)-$last_cut>200);
      }
      $string=~s/^\s*\+//;
      $string.="=0";
      print "R305_$equations_no: $string\n";
      check_equation($string);
      $equations_no++;
    }
  }
}
```

```
#3.6 in new version
for (my $i=1;$i<=$M;$i++)
{
  next unless (check_any_x($i,'*','*'));
  for (my $j=1;$j<=$M;$j++)
  {
    next unless (check_any_x($i,'*',$j));
    next if ($i==$j);
    for (my $r=1;$r<=$R-4+1;$r++)
    {
      next unless ((check_any_x($i,$r,$j))&&(check_any_x($j,$r+1,'*')));
      for (my $t=1;$t<=$M;$t++)
      {
        next if (($t==$i)||($t==$j));
        next unless (check_any_x($j,$r+1,$t));
        $string=add_variable('+',1,'y',$i,$r,$j,$j,$r+1,$t);
        $last_cut=0;
        for (my $k=1;$k<=$M;$k++)
        {
          next if (($k==$i)||($k==$j)||($k==$t));
          next unless (check_any_x($t,$r+2,$k));
          $string.=add_variable('-',1,'y',$i,$r,$j,$t,$r+2,$k);
          $last_cut=length($string), $string.="\n"  if (length($string)-$last_cut>200);
        }
        $string=~s/^\s*\+//;
        $string.="=0";
        print "R306_$equations_no: $string\n";
        check_equation($string);
        $equations_no++;
      }
    }
  }
}

#3.7 in new version
for (my $i=1;$i<=$M;$i++)
{
  #print "R7 i:$i\n";
  next unless (check_any_x($i,'*','*'));
  for (my $j=1;$j<=$M;$j++)
  {
    #print "  R7 j:$j\n";
    next if ($i==$j);
    next unless (check_any_x($i,'*',$j));
    for (my $t=1;$t<=$M;$t++)
    {
      #print "    R7 t:$t\n";
      next if (($t==$i)||($t==$j));
      next unless ((check_any_x($t,'*','*'))||(check_any_x('*','*',$t)));
      for (my $r=1;$r<=$R-5+1;$r++)
      {
        #print "    R7 s:$r\n";
        next unless (check_any_x($i,$r,$j));
        for (my $s=$r+2;$s<=$R-3+1;$s++)
        {
          print "      R7 r:$s ($i,$r,$j,*,$s,$t)\n" if ($debug);
          next unless ((check_any_x($t,$s+1,'*'))||(check_any_x('*',$s,$t)));
          $string='';
          $last_cut=0;
          for (my $k=1;$k<=$M;$k++)
          {
            print "        R7 k:$k (i:$i,s:$r,j:$j,t:$t,r:$s+1,k:$k)\n"  if ($debug);
            next if ((($k==$i)&&(($r>1)||($s<$R-1)))||($k==$j)||($k==$t));
            next unless ((check_any_x($t,$s+1,$k))||(check_any_x($k,$s,$t)));
            $string.=add_variable('+',1,'y',$i,$r,$j,$k,$s,$t);
            $string.=add_variable('-',1,'y',$i,$r,$j,$t,$s+1,$k);
            $last_cut=length($string), $string.="\n"  if (length($string)-$last_cut>200);
          }
          if ($string)
          {
            $string=~s/^\s*\+//;
            $string.="=0";
            print "R307_$equations_no: $string\n";
            check_equation($string);
            $equations_no++;
            die if (($debug)&&(length($string)<20));
          }
```

```
        }
      }
    }
  }
}

#3.8 in new version
for (my $i=1;$i<=$M;$i++)
{
  next unless (check_any_x($i,'*','*'));
  for (my $u=1;$u<=$M;$u++)
  {
    next if ($u==$i);
    next unless ((check_any_x($i,'*',$u))&&(check_any_x($u,'*','*')));
    for (my $v=1;$v<=$M;$v++)
    {
      next if (($v==$i)||($v==$u));
      next unless (check_any_x($u,'*',$v));
      for (my $p=2;$p<=$R-3+1;$p++)
      {
        next unless ((check_any_x($i,$p-1,$u))&&(check_any_x($u,$p,$v))&&(check_any_y($i,$p-
1,$u,$u,$p,$v)));
        $string=add_variable('+',1,'y',$i,$p-1,$u,$u,$p,$v);
        $last_cut=0;
        for (my $t=1;$t<=$M;$t++)
        {
          next if (($t==$i)||($t==$u)||($t==$v));
          next unless (check_any_x($v,$p+1,$t));
          $string.=add_variable('-',1,'z',$i,$p-1,$u,$u,$p,$v,$v,$p+1,$t);
          $last_cut=length($string), $string.="\n"  if (length($string)-$last_cut>200);
        }
        $string=~s/^\s*\+//;
        $string.="=0";
        print "R308_$equations_no: $string\n";
        check_equation($string);
        $equations_no++;
      }
    }
  }
}

#3.9 in new version
for (my $i=1;$i<=$M;$i++)
{
  next unless (check_any_x($i,'*','*'));
  for (my $u=1;$u<=$M;$u++)
  {
    next if ($u==$i);
    next unless ((check_any_x($i,'*',$u))&&(check_any_x($u,'*','*')));
    for (my $v=1;$v<=$M;$v++)
    {
      next if (($v==$i)||($v==$u));
      next unless (check_any_x($u,'*',$v));
      for (my $p=2;$p<=$R-3+1;$p++)
      {
        next unless ((check_any_x($i,$p-1,$u))&&(check_any_x($u,$p,$v))&&(check_any_y($i,$p-
1,$u,$u,$p,$v)));
        for (my $s=$p+2;$s<=$R;$s++)
        {
          $string=add_variable('+',1,'y',$i,$p-1,$u,$u,$p,$v);
          $last_cut=0;
          for (my $k=1;$k<=$M;$k++)
          {
            next if (($k==$i)||($k==$u)||($k==$v));
            next unless (check_any_x($k,$s,'*'));
            for (my $t=1;$t<=$M;$t++)
            {
              next if ((($t==$i)&&(($s<$R)||($p>2)))||($t==$u)||($t==$v)||($t==$k));
              next unless (check_any_x($k,$s,$t));
              $string.=add_variable('-',1,'z',$i,$p-1,$u,$u,$p,$v,$k,$s,$t);
              $last_cut=length($string), $string.="\n"  if (length($string)-$last_cut>200);
            }
          }
          $string=~s/^\s*\+//;
          $string.="=0";
          print "R309_$equations_no: $string\n";
          check_equation($string);
```

```
          $equations_no++;
        }
      }
    }
  }
}

#3.10 in new version
for (my $i=1;$i<=$M;$i++)
{
  next unless (check_any_x($i,'*','*'));
  for (my $j=1;$j<=$M;$j++)
  {
    next if ($j==$i);
    next unless (check_any_x($i,'*',$j));
    for (my $u=1;$u<=$M;$u++)
    {
      next if (($u==$i)||($u==$j));
      next unless (check_any_x($u,'*','*'));
      for (my $v=1;$v<=$M;$v++)
      {
        next if (($v==$i)||($v==$j)||($v==$u));
        next unless (check_any_x($u,'*',$v));
        for (my $p=3;$p<=$R-3+1;$p++)
        {
          next unless (check_any_x($u,$p,$v));
          for (my $r=1;$r<=$p-2;$r++)
          {
            next unless (check_any_x($i,$r,$j)&&(check_any_y($i,$r,$j,$u,$p,$v)));
            $string=add_variable('+',1,'y',$i,$r,$j,$u,$p,$v);
            $last_cut=0;
            for (my $t=1;$t<=$M;$t++)
            {
              next if ((($t==$i)&&(($r>1)||($p<$R-1)))||($t==$j)||($t==$u)||($t==$v));
              next unless (check_any_x($v,$p+1,$t));
              $string.=add_variable('-',1,'z',$i,$r,$j,$u,$p,$v,$v,$p+1,$t);
              $last_cut=length($string), $string.="\n"  if (length($string)-$last_cut>200);
            }
            $string=~s/^\s*\+//;
            $string.="=0";
            print "R310_$equations_no: $string\n";
            check_equation($string);
            $equations_no++;
          }
        }
      }
    }
  }
}

#3.11 in new version
for (my $i=1;$i<=$M;$i++)
{
  next unless (check_any_x($i,'*','*'));
  for (my $j=1;$j<=$M;$j++)
  {
    next if ($j==$i);
    next unless (check_any_x($i,'*',$j));
    for (my $u=1;$u<=$M;$u++)
    {
      next if (($u==$i)||($u==$j));
      next unless (check_any_x($u,'*','*'));
      for (my $v=1;$v<=$M;$v++)
      {
        next if (($v==$i)||($v==$j)||($v==$u));
        next unless (check_any_x($u,'*',$v));
        for (my $p=3;$p<=$R-4+1;$p++)
        {
          next unless (check_any_x($u,$p,$v));
          for (my $r=1;$r<=$p-2;$r++)
          {
            next unless (check_any_x($i,$r,$j)&&(check_any_y($i,$r,$j,$u,$p,$v)));
            for (my $s=$p+2;$s<=$R;$s++)
            {
              $string=add_variable('+',1,'y',$i,$r,$j,$u,$p,$v);
              $last_cut=0;
              for (my $k=1;$k<=$M;$k++)
```

```
              {
                next if (($k==$i)||($k==$j)||($k==$u)||($k==$v));
                next unless (check_any_x($k,$s,'*'));
                for (my $t=1;$t<=$M;$t++)
                {
                  next if ((($t==$i)&&(($s<$R)||($r>1)))||($t==$j)||($t==$u)||($t==$v)||($t==$k));
                  next unless (check_any_x($k,$s,$t));
                  $string.=add_variable('-',1,'z',$i,$r,$j,$u,$p,$v,$k,$s,$t);
                  $last_cut=length($string), $string.="\n"  if (length($string)-$last_cut>200);
                }
              }
              $string=~s/^\s*\+//;
              $string.="=0";
              print "R311_$equations_no: $string\n";
              check_equation($string);
              $equations_no++;
            }
          }
        }
      }
    }
  }
}

#3.12 in new version
for (my $i=1;$i<=$M;$i++)
{
  next unless (check_any_x($i,'*','*'));
  for (my $j=1;$j<=$M;$j++)
  {
    next if ($j==$i);
    next unless (check_any_x($i,'*',$j));
    for (my $k=1;$k<=$M;$k++)
    {
      next if (($k==$i)||($k==$j));
      next unless (check_any_x($k,'*','*'));
      for (my $t=1;$t<=$M;$t++)
      {
        next if (($t==$i)||($t==$j)||($t==$k));
        next unless (check_any_x($k,'*',$t));
        for (my $r=1;$r<=$R-4+1;$r++)
        {
          next unless
((check_any_x($i,$r,$j))&&(check_any_x($k,$r+2,$t))&&(check_any_y($i,$r,$j,$k,$r+2,$t)));
          $string=add_variable('+',1,'y',$i,$r,$j,$k,$r+2,$t);
          $string.=add_variable('-',1,'z',$i,$r,$j,$j,$r+1,$k,$k,$r+2,$t);
          $string=~s/^\s*\+//;
          $string.="=0";
          if (length($string)>3)
          {
            print "R312_$equations_no: $string\n";
            check_equation($string);
            $equations_no++;
          }
        }
      }
    }
  }
}

#3.13 in new version
for (my $i=1;$i<=$M;$i++)
{
  next unless (check_any_x($i,'*','*'));
  for (my $j=1;$j<=$M;$j++)
  {
    next if ($j==$i);
    next unless (check_any_x($i,'*',$j));
    for (my $k=1;$k<=$M;$k++)
    {
      next if (($k==$i)||($k==$j));
      next unless (check_any_x($k,'*','*'));
      for (my $t=1;$t<=$M;$t++)
      {
        next if (($t==$j)||($t==$k));
        next unless (check_any_x($k,'*',$t));
        for (my $r=1;$r<=$R-5+1;$r++)
```

```
        {
          next unless (check_any_x($i,$r,$j));
          for (my $s=$r+3;$s<=$R;$s++)
          {
            next if (($t==$i)&&(($r>1)||($s<$R)));
            next unless ((check_any_x($k,$s,$t))&&(check_any_y($i,$r,$j,$k,$s,$t)));
            $string=add_variable('+',1,'y',$i,$r,$j,$k,$s,$t);
            $last_cut=0;
            for (my $v=1;$v<=$M;$v++)
            {
              next if (($v==$i)||($v==$j)||($v==$k)||($v==$t));
              next unless (check_any_x($j,$r+1,$v));
              $string.=add_variable('-',1,'z',$i,$r,$j,$j,$r+1,$v,$k,$s,$t);
              $last_cut=length($string), $string.="\n"  if (length($string)-$last_cut>200);
            }
            $string=~s/^\s*\+//;
            $string.="=0";
            if (length($string)>3)
            {
              print "R313_$equations_no: $string\n";
              check_equation($string);
              $equations_no++;
            }
          }
        }
      }
    }
  }
}

#3.14 in new version
for (my $i=1;$i<=$M;$i++)
{
  next unless (check_any_x($i,'*','*'));
  for (my $j=1;$j<=$M;$j++)
  {
    next if ($j==$i);
    next unless (check_any_x($i,'*',$j));
    for (my $k=1;$k<=$M;$k++)
    {
      next if (($k==$i)||($k==$j));
      next unless (check_any_x($k,'*','*'));
      for (my $t=1;$t<=$M;$t++)
      {
        next if (($t==$j)||($t==$k));
        next unless (check_any_x($k,'*',$t));
        for (my $r=1;$r<=$R-5+1;$r++)
        {
          next unless (check_any_x($i,$r,$j));
          for (my $s=$r+3;$s<=$R;$s++)
          {
            next if (($i==$t)&&(($r>1)||($s<$R)));
            next unless ((check_any_x($k,$s,$t))&&(check_any_y($i,$r,$j,$k,$s,$t)));
            $string=add_variable('+',1,'y',$i,$r,$j,$k,$s,$t);
            $last_cut=0;
            for (my $u=1;$u<=$M;$u++)
            {
              next if (($u==$i)||($u==$j)||($u==$k)||($u==$t));
              $string.=add_variable('-',1,'z',$i,$r,$j,$u,$s-1,$k,$k,$s,$t);
              $last_cut=length($string), $string.="\n"  if (length($string)-$last_cut>200);
            }
            $string=~s/^\s*\+//;
            $string.="=0";
            if (length($string)>3)
            {
              print "R314_$equations_no: $string\n";
              check_equation($string);
              $equations_no++;
            }
          }
        }
      }
    }
  }
}

#3.15 in new version
```

```
for (my $i=1;$i<=$M;$i++)
{
  next unless (check_any_x($i,'*','*'));
  for (my $j=1;$j<=$M;$j++)
  {
    next if ($j==$i);
    next unless (check_any_x($i,'*',$j));
    for (my $k=1;$k<=$M;$k++)
    {
      next if (($k==$i)||($k==$j));
      next unless (check_any_x($k,'*','*'));
      for (my $t=1;$t<=$M;$t++)
      {
        next if (($t==$j)||($t==$k));
        next unless (check_any_x($k,'*',$t));
        for (my $r=1;$r<=$R-6+1;$r++)
        {
          next unless (check_any_x($i,$r,$j));
          for (my $s=$r+4;$s<=$R;$s++)
          {
            next unless (check_any_x($k,$s,$t));
            for (my $p=$r+2;$p<=$s-2;$p++)
            {
              next if (($i==$t)&&(($r>1)||($s<$R)));
              next unless (check_any_y($i,$r,$j,$k,$s,$t));
              $string=add_variable('+',1,'y',$i,$r,$j,$k,$s,$t);
              $last_cut=0;
              for (my $u=1;$u<=$M;$u++)
              {
                next if (($u==$i)||($u==$j)||($u==$k)||($u==$t));
                next unless (check_any_x($u,$p,'*'));
                for (my $v=1;$v<=$M;$v++)
                {
                  next if (($v==$i)||($v==$j)||($v==$k)||($v==$t)||($v==$u));
                  next unless (check_any_x($u,$p,$v));
                  $string.=add_variable('-',1,'z',$i,$r,$j,$u,$p,$v,$k,$s,$t);
                  $last_cut=length($string), $string.="\n"  if (length($string)-$last_cut>200);
                }
              }
              $string=~s/^\s*\+//;
              $string.="=0";
              if (length($string)>3)
              {
                print "R315_$equations_no: $string\n";
                check_equation($string);
                $equations_no++;
              }
            }
          }
        }
      }
    }
  }
}

#3.16 in new version
for (my $u=1;$u<=$M;$u++)
{
  next unless (check_any_x($u,'*','*'));
  for (my $v=1;$v<=$M;$v++)
  {
    next if ($u==$v);
    next unless (check_any_x($u,'*',$v));
    for (my $t=1;$t<=$M;$t++)
    {
      next if (($t==$u)||($t==$v));
      next unless (check_any_x($v,'*',$t));
      for (my $p=2;$p<=$R-3+1;$p++)
      {
        next unless (check_any_y($u,$p,$v,$v,$p+1,$t));
        $string=add_variable('+',1,'y',$u,$p,$v,$v,$p+1,$t);
        $last_cut=0;
        for (my $i=1;$i<=$M;$i++)
        {
          next if (($i==$u)||($i==$v)||($i==$t));
          $string.=add_variable('-',1,'z',$i,$p-1,$u,$u,$p,$v,$v,$p+1,$t);
          $last_cut=length($string), $string.="\n"  if (length($string)-$last_cut>200);
```

```
          }
          $string=~s/^\s*\+//;
          $string.="=0";
          print "R316_$equations_no: $string\n";
          check_equation($string);
          $equations_no++;
        }
      }
    }
  }

#3.17 in new version
for (my $u=1;$u<=$M;$u++)
{
  next unless (check_any_x($u,'*','*'));
  for (my $v=1;$v<=$M;$v++)
  {
    next if ($u==$v);
    next unless (check_any_x($u,'*',$v));
    for (my $t=1;$t<=$M;$t++)
    {
      next if (($t==$u)||($t==$v));
      next unless (check_any_x($v,'*',$t));
      for (my $p=3;$p<=$R-3+1;$p++)
      {
        next unless (check_any_y($u,$p,$v,$v,$p+1,$t));
        for (my $r=1;$r<=$p-2;$r++)
        {
          $string=add_variable('+',1,'y',$u,$p,$v,$v,$p+1,$t);
          $last_cut=0;
          for (my $i=1;$i<=$M;$i++)
          {
            next if (($i==$u)||($i==$v)||(($i==$t)&&(($r>1)||($p<$R-1))));
            next unless (check_any_x($i,$r,'*'));
            for (my $j=1;$j<=$M;$j++)
            {
              next if (($j==$u)||($j==$v)||($j==$t)||($j==$i));
              next unless (check_any_x($i,$r,$j));
              $string.=add_variable('-',1,'z',$i,$r,$j,$u,$p,$v,$v,$p+1,$t);
              $last_cut=length($string), $string.="\n"  if (length($string)-$last_cut>200);
            }
          }
          $string=~s/^\s*\+//;
          $string.="=0";
          print "R317_$equations_no: $string\n";
          check_equation($string);
          $equations_no++;
        }
      }
    }
  }
}

#3.18 in new version
for (my $u=1;$u<=$M;$u++)
{
  next unless (check_any_x($u,'*','*'));
  for (my $v=1;$v<=$M;$v++)
  {
    next if ($u==$v);
    next unless (check_any_x($u,'*',$v));
    for (my $k=1;$k<=$M;$k++)
    {
      next if (($k==$u)||($k==$v));
      next unless (check_any_x($k,'*','*'));
      for (my $t=1;$t<=$M;$t++)
      {
        next if (($t==$u)||($t==$v)||($t==$k));
        next unless (check_any_x($k,'*',$t));
        for (my $p=2;$p<=$R-4+1;$p++)
        {
          next unless (check_any_x($u,$p,$v));
          for (my $s=$p+2;$s<=$R;$s++)
          {
            next unless ((check_any_x($k,$s,$t))&&(check_any_y($u,$p,$v,$k,$s,$t)));
            $string=add_variable('+',1,'y',$u,$p,$v,$k,$s,$t);
            $last_cut=0;
```

```
            for (my $i=1;$i<=$M;$i++)
            {
              next if (($i==$u)||($i==$v)||($i==$k)||(($i==$t)&&(($s<$R)||($p>2))));
              next unless (check_any_x($i,$p-1,$u));
              $string.=add_variable('-',1,'z',$i,$p-1,$u,$u,$p,$v,$k,$s,$t);
              $last_cut=length($string), $string.="\n"  if (length($string)-$last_cut>200);
            }
            $string=~s/^\s*\+//;
            $string.="=0";
            print "R318_$equations_no: $string\n";
            check_equation($string);
            $equations_no++;
          }
        }
      }
    }
  }
}

#3.19
for (my $u=1;$u<=$M;$u++)
{
  next unless (check_any_x($u,'*','*'));
  for (my $v=1;$v<=$M;$v++)
  {
    next if ($u==$v);
    next unless (check_any_x($u,'*',$v));
    for (my $k=1;$k<=$M;$k++)
    {
      next if (($k==$u)||($k==$v));
      next unless (check_any_x($k,'*','*'));
      for (my $t=1;$t<=$M;$t++)
      {
        next if (($t==$u)||($t==$v)||($t==$k));
        next unless (check_any_x($k,'*',$t));
        for (my $p=3;$p<=$R-4+1;$p++)
        {
          next unless (check_any_x($u,$p,$v));
          for (my $r=1;$r<=$p-2;$r++)
          {
            for (my $s=$p+2;$s<=$R;$s++)
            {
              next unless ((check_any_x($k,$s,$t))&&(check_any_y($u,$p,$v,$k,$s,$t)));
              $string=add_variable('+',1,'y',$u,$p,$v,$k,$s,$t);
              $last_cut=0;
              for (my $i=1;$i<=$M;$i++)
              {
                next if (($i==$u)||($i==$v)||($i==$k)||(($i==$t)&&(($s<$R)||($r>1))));
                next unless (check_any_x($i,$r,'*'));
                for (my $j=1;$j<=$M;$j++)
                {
                  next if (($j==$u)||($j==$v)||($j==$k)||($j==$t)||($j==$i));
                  next unless (check_any_x($i,$r,$j));
                  $string.=add_variable('-',1,'z',$i,$r,$j,$u,$p,$v,$k,$s,$t);
                  $last_cut=length($string), $string.="\n"  if (length($string)-$last_cut>200);
                }
              }
              $string=~s/^\s*\+//;
              $string.="=0";
              print "R319_$equations_no: $string\n";
              check_equation($string);
              $equations_no++;
            }
          }
        }
      }
    }
  }
}

die if ($debug);
print "Bounds\n";
foreach (sort keys %variables)
{
  my $key=$_;
  my @arr=split(/\_/,$key);
  if ($restrict_boundaries)
```

```
  {
    print "".($solution{$key})." <= $key <= ".($solution{$key})."\n";
  }
  else
  {
    print "".(0)." <= $key <= ".($total_flow_constant)."\n";
  }
}
print "End\n";
```